\documentclass{jfm}
\usepackage{graphicx}
\usepackage{epstopdf, epsfig}
\usepackage[caption=false]{subfig}
\usepackage{tikz} 
\usepackage{pgfplots}
\usepackage{natbib}
\usepackage{amsmath}
\usepackage{multirow}
\pgfplotsset{compat=newest}
\pgfplotsset{scaled y ticks=false}
\pgfplotsset{scaled x ticks=false}
\usepackage[normalem]{ulem}

\definecolor{mycolor1}{rgb}{0.07843,0.07843,0.07843}% black
\definecolor{mycolor2}{rgb}{0.216 0.5 0.72}%Blue
\definecolor{mycolor3}{rgb}{0.89 0.1 0.11}% Red
\definecolor{mycolor4}{rgb}{0.3 0.69 0.29}% green
\definecolor{mycolor5}{rgb}{0.56863,0.43137,0.86275}%purple
\definecolor{darkgreen}{RGB}{0,102,51}%
\definecolor{mycolor22}{rgb}{0.19608,0.72549,0.62745}%GreenRe
\definecolor{mycolor33}{rgb}{0.21569,0.49412,0.86275}%BlueRe

\newcommand{\blackthin}{\raisebox{2pt}{\tikz{\draw[-,mycolor1,solid,line width = 0.5pt](0,0) -- (5mm,0);}}}
\newcommand{\redline}{\raisebox{2pt}{\tikz{\draw[-,red, line width = 0.75pt](0,0) -- (5mm,0);}}}
\newcommand{\reddashed}{\raisebox{2pt}{\tikz{\draw[-,red,dashed,line width = 0.75pt](0,0) -- (5mm,0);}}}
\newcommand{\reddotted}{\raisebox{2pt}{\tikz{\draw[-,red,dotted,line width = 0.75pt](0,0) -- (5mm,0);}}}
\newcommand{\violetline}{\raisebox{2pt}{\tikz{\draw[-,violet, line width = 0.75pt](0,0) -- (5mm,0);}}}
\newcommand{\violetdashed}{\raisebox{2pt}{\tikz{\draw[-,violet,dashed,line width = 0.75pt](0,0) -- (5mm,0);}}}
\newcommand{\blueline}{\raisebox{2pt}{\tikz{\draw[-,blue, line width = 0.75pt](0,0) -- (5mm,0);}}}
\newcommand{\bluedashed}{\raisebox{2pt}{\tikz{\draw[-,blue,dashed,line width = 0.75pt](0,0) -- (5mm,0);}}}

\newcommand{\newredline}{\raisebox{2pt}{\tikz{\draw[-,mycolor3,line width = 0.75pt](0,0) -- (5mm,0);}}}

\newcommand{\darkgreenline}{\raisebox{2pt}{\tikz{\draw[-,darkgreen, line width = 0.75pt](0,0) -- (5mm,0);}}}
\newcommand{\darkgreendashed}{\raisebox{2pt}{\tikz{\draw[-,darkgreen,dashed,line width = 0.75pt](0,0) -- (5mm,0);}}}

\newcommand{\blueReline}{\raisebox{2pt}{\tikz{\draw[-,mycolor33, line width = 0.75pt](0,0) -- (5mm,0);}}}
\newcommand{\mylab}[3]{\raisebox{#2}[0mm][0mm]{\makebox[0mm][l]{\hspace*{#1}#3}}}

\shorttitle{Scaling and dynamics of turbulence over sparse canopies}
\shortauthor{A. Sharma and R. Garc{\'i}a-Mayoral}

\title{Scaling and dynamics of turbulence over sparse canopies}

\author{Akshath Sharma\aff{1}
\and Ricardo Garc{\'i}a-Mayoral\aff{1}
  \corresp{\email{r.gmayoral@eng.cam.ac.uk}}
 }

\affiliation{\aff{1}Department of Engineering, University of Cambridge,
Trumpington Street, Cambridge CB2~1PZ, UK
}

\begin{document}
\maketitle
\begin{abstract}

Turbulent flows within and over sparse canopies are investigated using direct numerical simulations {at moderate friction Reynolds numbers $\Rey_\tau \approx 520$ and $1000$. The height of the canopies studied is $h^+ \approx 110$--$200$, which is typical of some engineering canopies but much lower than for most vegetation canopies. The analysis of the effect of Reynolds number in our simulations, however, suggests that the dynamics observed would be relevant for larger Reynolds numbers as well.} {In channel flows, the distribution of the total stress is linear with height. Over smooth walls, the total stress is the sum of the viscous and the Reynolds shear stresses, the `fluid stress' $\tau_f$. In canopies, in turn, there is an additional contribution from the canopy drag, which can dominate within.} {Furthermore, the full Reynolds shear stress has contributions from the dispersive, element-induced flow and from the background turbulence, the part of the flow that remains once the element-induced flow is filtered out.} {For {the present} sparse canopies, we find that the {ratio} of the viscous stress and the {background} Reynolds shear stress to {their sum,} $\tau_f$, is similar to that over smooth-walls at each height, even within the canopy.} From this, a height-dependent scaling based on $\tau_f$ is proposed. Using this scaling, the background turbulence within the canopy shows similarities with turbulence over smooth walls. This suggests that the background turbulence scales with $\tau_f$, rather than with the conventional scaling based on the total stress. This effect is essentially captured when the canopy is substituted by a drag force that acts on the mean velocity profile alone, aiming to produce the correct $\tau_f$, {without the discrete presence of the canopy elements acting directly on the fluctuations.} The proposed mean-only forcing is shown to produce better estimates for the turbulent fluctuations compared to a conventional, homogeneous-drag model. {These results suggest that a sparse canopy acts on the background turbulence primarily through the change it induces on the mean velocity profile, which in turn sets the scale for turbulence, rather than through a direct interaction of the canopy elements with the fluctuations.} The effect of the element-induced flow, however, requires the representation of the individual canopy elements.

\end{abstract}


\section{Introduction} \label{sec:intro}

Canopy flows are ubiquitous in both natural and artificial settings. {Although they have mostly been studied in the framework of flows through crops and forests \citep{Finnigan2000,Belcher2012,Nepf2012}, they are also relevant to flows over engineered surfaces, such as pin fins for heat transfer or piezoelectric filaments for energy harvesting \citep{Bejan1993,McCloskey2017}. While the latter usually have small to moderate Reynolds numbers, vegetation canopies typically have much larger ones.} The study of turbulent flows over canopies has wide ranging applications, including reducing crop loss \citep{deLangre2008}, energy harvesting \citep{Mcgarry2011,McCloskey2017,Elahi2018} and improving heat transfer \citep{Fazu1989,Bejan1993}. 
On the basis of the geometry and spacing of the canopy elements, a canopy can be classified as dense, sparse or transitional \citep{Nepf2012}.  In the dense limit, the canopy elements are in close proximity to each other and turbulence is essentially not able to penetrate within the canopy layer. In the sparse limit, the spacing between canopy elements is large and the turbulent eddies can penetrate the full depth of the canopy. An intermediate or transitional regime lies between these two limits. Turbulent flows in the dense regime, reviewed by \citet{Finnigan2000} and \citet{Nepf2012}, are characterised by the formation of Kelvin--Helmholtz-like, or mixing-layer, instabilities, originating from the inflection point at the canopy tips \citep{Raupach1996}. As the sparsity of the canopies is increased, the importance of the Kelvin--Helmholtz-like instability decreases \citep{Poggi2004,Pietri2009,Huang2009}. Eventually, the flow would resemble that over a smooth wall, albeit perturbed by the discrete presence of the individual canopy elements \citep{Finnigan2000}. The separation between these regimes is still somewhat unclear. \citet{Nepf2012} proposed an approximate classification of the canopy regime based on the roughness frontal density, $\lambda_f$. \citet{Nepf2012} observed that canopies are dense for $\lambda_f \gg 0.1$,  sparse for $\lambda_f \ll 0.1$, and intermediate for $\lambda_f \approx 0.1$. However, in addition to the geometric parameter $\lambda_f$, the lengthscales of the flow should also be considered when determining the regime of the canopy. The lengthscales in a turbulent flow may be much larger than the element spacing at a particular Reynolds number, so that the turbulent eddies are precluded from penetrating within the canopy. As the Reynolds number is increased, however, the turbulent lengthscales will eventually become comparable to the element spacing and allow the turbulent eddies to penetrate within the canopy efficiently.

In the present work, we study flows within and above sparse canopies {using direct numerical simulation (DNS). While this allows for the full resolution of turbulence, it restricts our simulations to moderate friction Reynolds numbers, $\Rey_\tau \approx 520$--$1000$, and element heights, $h^+ \approx 110$--$200$. While these heights would be directly applicable to some of the engineered canopies mentioned above, they are much smaller than those typical of vegetation canopies, $h^+ \approx 10^4$--$10^6$ \citep[e.g.,][]{Green1995,Novak2000,Zhu2006}, although comparable to some laboratory experiments $h^+ \approx 400$--$800$ \citep{Raupach2006,Bohm2013}. In any event, we provide evidence in \S\ref{subsec:re_effect} of the scaling of the canopy-flow dynamics with the Reynolds number, which would make our conclusions of relevance for canopies at larger Reynolds numbers as well.} The {present} canopies have low roughness densities $\lambda_f \lesssim 0.1$, with element spacings large enough to limit their interaction with the near-wall turbulence dynamics. Owing to the sparse nature of these canopies, we would expect the flow within them to be dominated by the footprint of the canopy elements, rather than by a mixing-layer instability. Conventionally, a homogeneous-drag is used to represent the effect of canopies \citep{Dupont2008,Finnigan2009,Huang2009,Bailey2016}. This approach would only be strictly valid to represent very closely packed canopies, where the element spacing is much smaller than any lengthscale in the overlying flow, and even small flow structures perceive the canopy elements as acting collectively \citep{Zampogna2016}. Using a homogeneous drag to capture the effects of sparser canopies tends to overdamp turbulent fluctuations within the canopies \citep{Yue2007,Bailey2013}. This is typically attributed to the inability of homogenised models to capture the element-induced flow, and the lack of representation of the gaps between the canopy elements, where the fluctuations would not experience any damping \citep{Bailey2013}. In the present work we separate the effect of the element-induced coherent flow from the incoherent background turbulence, and focus mainly on the properties of the latter. We study different element spacings and geometries. We propose a scaling that suggests that the dynamics of the background turbulence within sparse canopies are mainly governed by their effect on the mean velocity, rather than by the direct interaction of the canopy elements with the flow. Based on this scaling, we propose that the effect on the background turbulence is represented better by a drag acting on the mean flow alone than by a homogeneous drag. Partial results from some of the simulations can be found in \cite{Sharma2018a,Sharma2018}.

The paper is organised as follows. The numerical methods used and the canopy geometries simulated are described in \S\ref{sec:num_method}. The results of the canopy-resolving simulations and the scaling of turbulent fluctuations are discussed in \S\ref{sec:resolved_canopy}. The results obtained from simulations that substitute the canopy by a drag force are discussed in \S\ref{sec:results_models}. The conclusions are presented in \S\ref{sec:conclusions}. 
  
 \begin{figure}
		\centering
           % \vspace{}
%  		   	\includegraphics[width=1\textwidth,trim={0 1.75cm 0 1.75cm},clip]{Images_new/canopy_schematic.png}
 	      \includegraphics[width=1\textwidth,trim={0 1.75cm 0 1.75cm},clip]{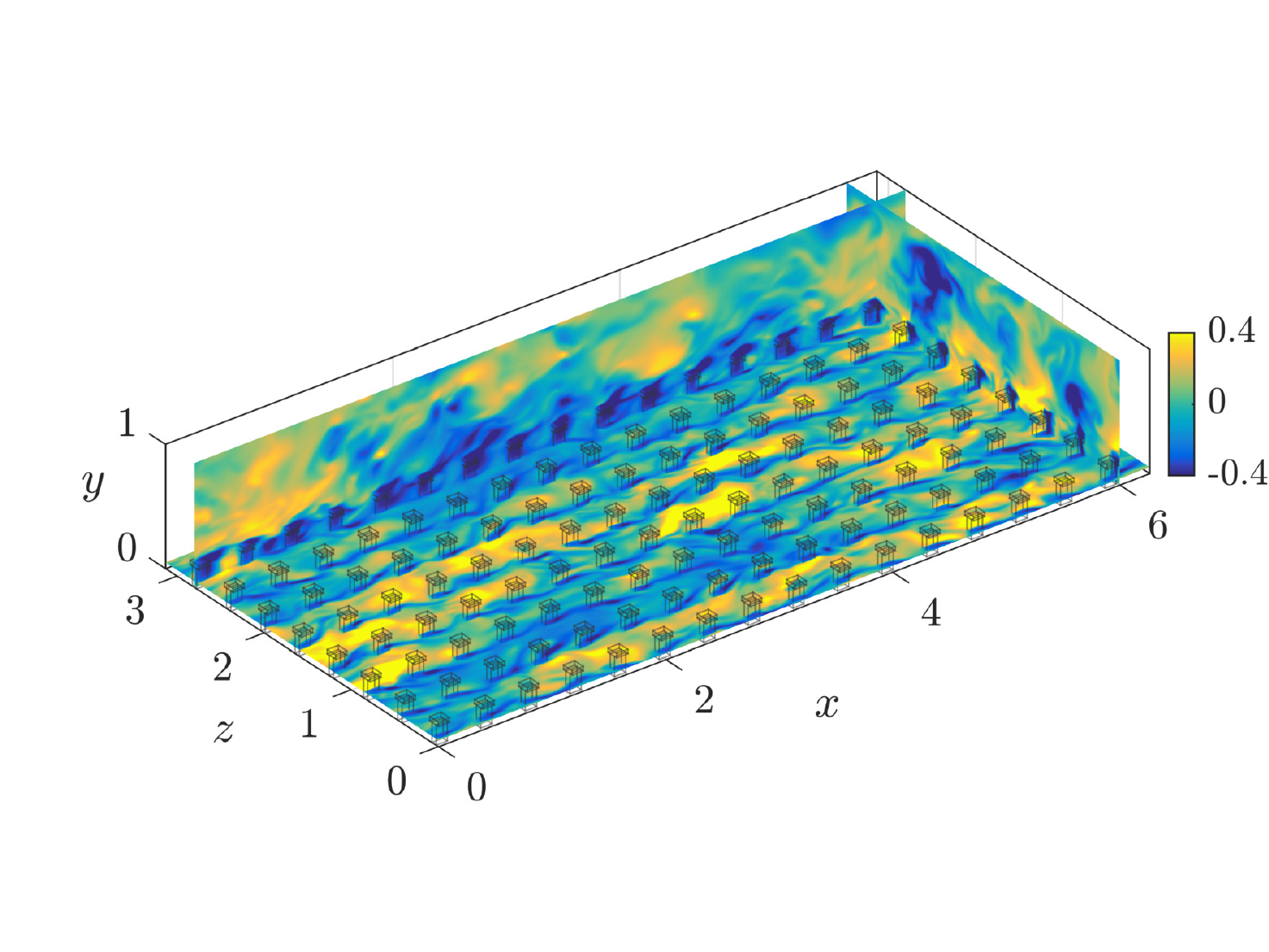}
  		    %\vspace{-1.5cm}
    		\caption{Schematic of the numerical domain. An instantaneous field of the fluctuating streamwise velocity, scaled , from case TP1 is shown in three orthogonal planes.}
	    	\label{fig:channel_schematic}	
 \end{figure}  
 
 \begin{table}
\begin{center}

%\lineup
\begin{tabular}{clccccccc}
%\br
           &Case               &$N_x \times N_z$          &{$u_\tau$}  &$\Rey_\tau$    & $\lambda_f$     & $\int D^+$    &$\Delta x^+$   &$\Delta z^+$  \\
\hline
%\mr
Smooth             &S                   &--                            &   {0.055}        &538.8            &--                   & --                    &8.8            &4.4 \\
\hline
                     &{PD}     & {64$\times$32}  & {0.153}     &{535.4}    &{0.88}      &{0.99}   &{4.38}   &{4.38}    \\
                     \\
                     &P0                  & 32$\times$16                   &   {0.182}       &532.5            &0.22                   &0.94              &4.36          &4.36          \\
              &P0-H              &--                                     &  {0.203}       &594.2             &--                      &0.93              &9.72          &4.86                \\
            &P0-H0            &--                                     &   {0.227}      &549.4              &--                      &0.90             &8.99          &4.49             \\
 \multirow{1}{*}{Impermeable}\\
 \multirow{1}{*}{prismatic}           &P1                   &16$\times$8                      &   {0.138}      &520.3             &0.05                  &0.79              &4.26          &4.26              \\
 \multirow{1}{*}{elements}           &P1-H0            &--                                      &   {0.147}     &553.8              &--                     & 0.80             &9.06          &4.53              \\
\\
            &P2                  &8$\times$4                        &   {0.093}     &529.4              &0.01                   &0.57             &4.33          &4.33              \\
            &P2-H0           &--                                      &   {0.092}    &522.9              &--                      &0.57                  &8.55          &4.28              \\
            \\
          &{P2I$_{Re}$}     &{16$\times$8}  &  {0.082}  &{1068.3}     &{0.01}    &{0.59}       &{4.37}  &{4.37}     \\
           &{P2O$_{Re}$}   &{8$\times$4}    &  {0.091}  &{1000.4}  &{0.01}    &{0.61}         &{4.09}   &{4.09}              \\
\hline
            &T1                  &16$\times$8                       &   {0.133}      &505.6               &0.07                   &0.80            &4.14         &4.14              \\
\multirow{2}{*}{Impermeable}            &T1-H              &--                                      &   {0.160}     &503.3                &--                     & 0.85           &11.00         &5.50              \\
\multirow{2}{*}{T-shaped}             &T1-H0           &--                                       &  {0.165}     &519.9                &--                      & 0.82           &11.34        &5.67              \\
\multirow{2}{*}{elements}    \\
         &T2                 &8$\times$4                         &   {0.090}    &513.3                 &0.02                 &0.58              &4.20          &4.20             \\
          &T2-H0           &--                                      &   {0.097}    &527.4                 &--                    &0.60              &8.63          &4.31               \\
\hline
\multirow{2}{*}{Permeable} \\
\multirow{2}{*}{T-shaped}             &TP1                &16$\times$8                 &   {0.160}     &505.5               &0.07                  &0.81            &8.27          &4.14              \\
\multirow{2}{*}{elements}          &TP1-L            &--                                      &   {0.167}      &527.1                &--                     &0.81              &8.62          &4.31             \\  
\hline
%\br
\end{tabular}
\caption{\label{tab:DNS_param} Simulation parameters. $N_x$ and $N_z$ are the number of canopy elements in the streamwise and spanwise directions, respectively, $u_\tau$ is the friction velocity based on the net drag {and scaled with the channel bulk velocity}, $\Rey_\tau$ is the friction Reynolds number based on $u_\tau$ and $\delta$, and $\lambda_f$ is the roughness frontal density. $\int D^+$ is the net canopy drag force scaled with $u_\tau$, that is, the proportion of the total drag on the fluid exerted by the canopy elements, with the remainder being the friction at the bottom wall. The grid resolutions in the streamwise and spanwise directions are $\Delta x^+$ and $\Delta z^+$, respectively.}
\end{center}
\end{table}   
 
\section{Numerical simulations} \label{sec:num_method}
We conduct direct numerical simulations of an open channel with canopy elements protruding from the wall. The streamwise, wall-normal and spanwise coordinates are $x$, $y$ and $z$ respectively, and the associated velocities $u$, $v$ and $w$. The size of the simulation domain is $2\pi \delta \times \delta \times \pi \delta$, with the channel height $\delta = 1$. This box size has been shown to be adequate to capture one-point statistics up to the channel height for the friction Reynolds numbers used in the present study \citep{Lozano2014}. A schematic representation of the numerical domain is shown in figure \ref{fig:channel_schematic}. The domain is periodic in the $x$ and $z$ directions. No-slip and impermeability conditions are applied at the bottom boundary, $y = 0$, and free slip and impermeability at the top, $y = \delta$. It is shown in \S\ref{sec:resolved_canopy} that the height of the roughness sublayer for the canopies studied here extends to only half of the domain height, so the top boundary of the channel does not interfere with the canopy flow. The flow is incompressible, with the density set to unity. All simulations are run at a constant mass flow rate, with the viscosity adjusted to obtain the desired friction Reynolds number based on the total stress. {Most simulations are conducted at a friction Reynolds number $\Rey_\tau = u_\tau \delta/\nu \approx 520$, and a few at $\Rey_\tau \approx 1000$.}

The numerical method used to solve the three-dimensional Navier-Stokes equations is adapted from \citet{Fairhall2018}. A Fourier spectral discretisation is used in the streamwise and spanwise directions. The wall-normal direction is discretised using a second-order centred difference scheme on a staggered grid. For the simulations at $\Rey_\tau \approx 520$, the grid in the wall-normal direction is stretched to give a resolution $\Delta y^+_{min} \approx 0.2$ at the wall, stretching to $\Delta y^+_{max} \approx 2$ at the top of the domain. {For the simulations at $\Rey_\tau \approx 1000$, wall-normal resolutions of $\Delta y^+_{min} \approx 0.35$ and $\Delta y^+_{max} \approx 5.5$ are used, as dissipation occurs at larger scales near the centre of the channel at larger Reynolds numbers \citep{Jimenez2012}.} The wall-parallel resolutions for the different cases are given in table~\ref{tab:DNS_param}. The time advancement is carried out using a three-step Runge-Kutta method with a fractional step, pressure correction method that enforces continuity \citep{Le1991}
\begin{eqnarray}
\label{eq:disc_NS}\left[\mathrm{I} - \Delta t \frac{\beta_k}{\Rey} \mathrm{L}\right] \boldsymbol{u}^n_k & = & \boldsymbol{u}^n_{k-1} + \Delta t \left[ \frac{\alpha_{k}}{\Rey}\mathrm{L} \boldsymbol{u}^n_{k-1} - \gamma_{k}\mathrm{N}(\boldsymbol{u}^n_{k-1}) - \right.\nonumber\\
&&\left. \zeta_k \mathrm{N}(\boldsymbol{u}^n_{k-2}) - (\alpha_k + \beta_k)\mathrm{G}(p^n_k) \vphantom{ \frac{\alpha_{k}}{\Rey}}\right], k\in[1,3],\\ 
\mathrm{DG}(\phi^n_k) & = &  \frac{1}{(\alpha_k + \beta_k) \Delta t} \mathrm{D}(\boldsymbol{u}^n_k), \\
\boldsymbol{u}^n_{k+1} & = &  \boldsymbol{u}^n_k - (\alpha_k + \beta_k)\Delta t G(\phi^n_k),\\
p^n_{k+1} & = &  p^n_k + \phi^n_k.
\end{eqnarray}
Where $\mathrm{I}$ is the identity matrix and L, D and G are the Laplacian, divergence and gradient operators respectively. N is the dealiased advective term. $\alpha_k$, $\beta_k$, $\gamma_k$ and $\zeta_k$ are the Runge-Kutta coefficients for substep $k$ from \citet{Le1991}, and $\Delta t$ is the timestep. 

\subsection{Canopy-resolving simulations}
\begin{figure}
		\centering
		\subfloat{%
  		\includegraphics[scale=0.8,trim={1.2cm 0cm 2cm 0cm},clip]{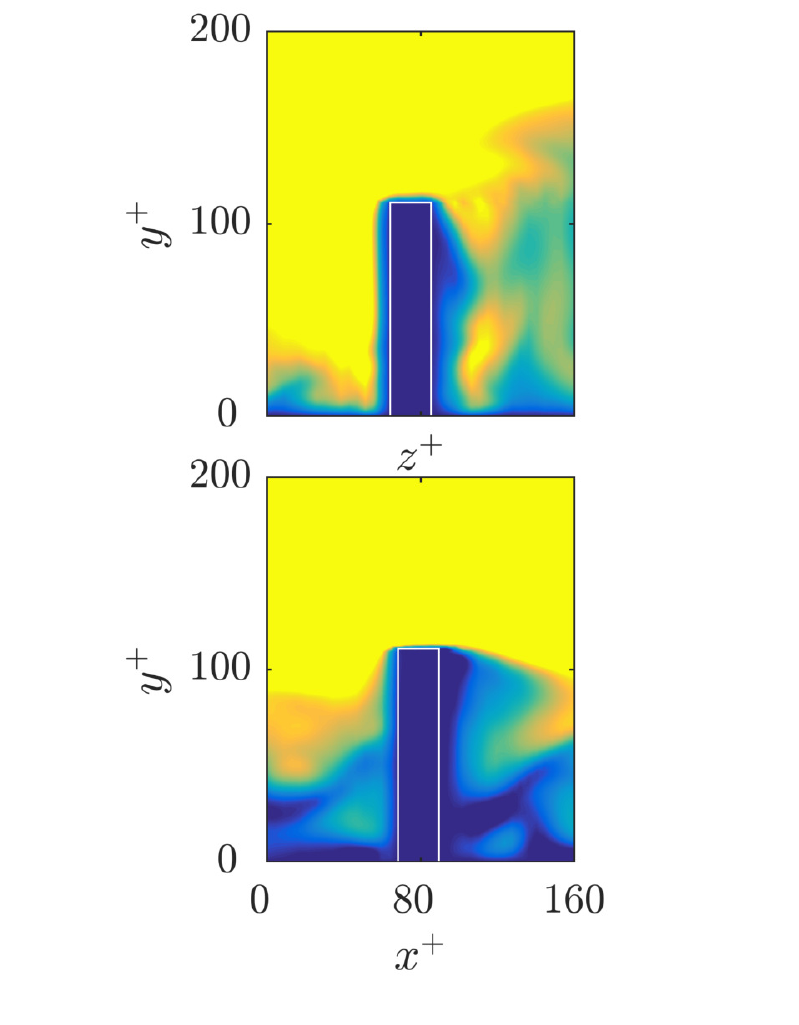}
  		}%
  		 \mylab{-2.8cm}{-0.55cm}{(\textit{a})}%
  		 \mylab{-2.8cm}{-4.2cm}{(\textit{d})}%
  		 \subfloat{%
       \includegraphics[scale=0.8,trim={2.5cm 0  2cm -0.2cm},clip]{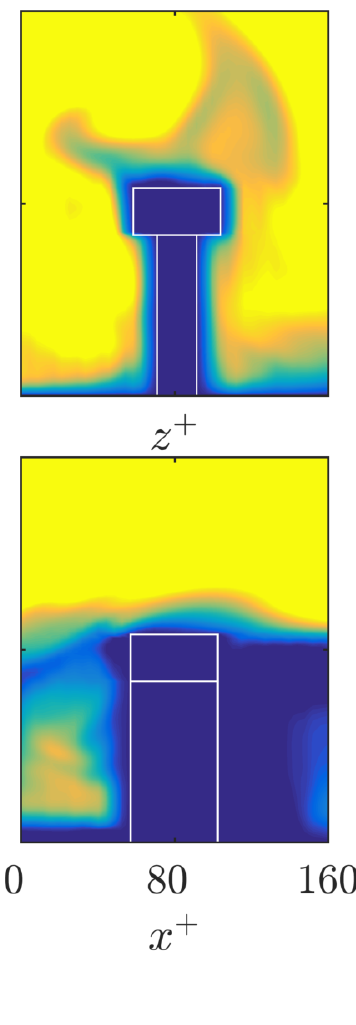}
  		}%
  		 \mylab{-2.8cm}{-0.55cm}{(\textit{b})}%
  		 \mylab{-2.8cm}{-4.2cm}{(\textit{e})}%
  		  		 \subfloat{%
         \includegraphics[scale=0.8,trim={2.5cm 0 0 -0.2cm},clip]{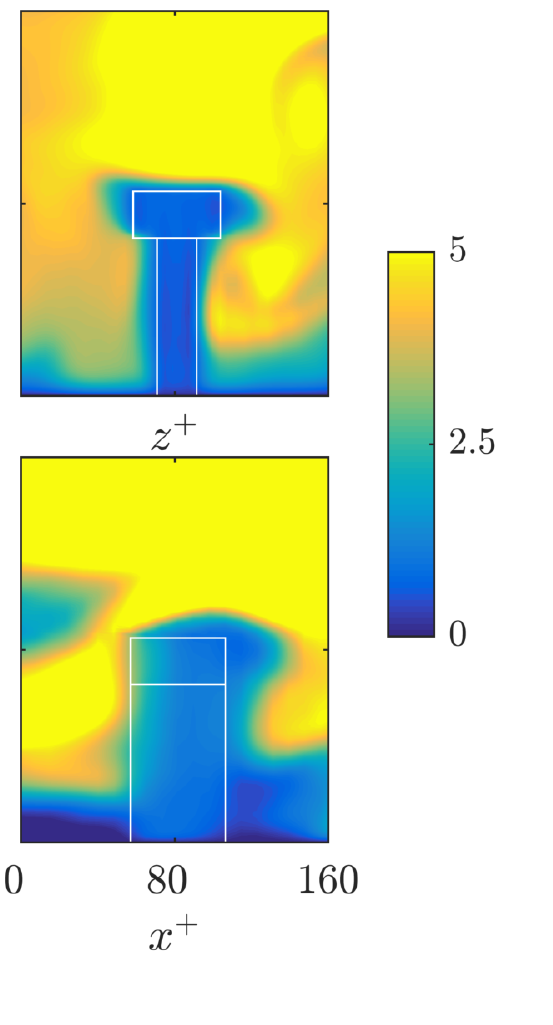}
  		}%
  		 \mylab{-4.4cm}{-0.55cm}{(\textit{c})}%
  		  \mylab{-4.4cm}{-4.2cm}{(\textit{f})}%
  	    %\vspace{-1.5cm}
    		\caption{Contours of instantaneous streamwise velocity in planes passing through the centre of a canopy element. ($a$--$c$) represent cuts in the $z$--$y$ plane, and ($d$--$f$) represent cuts in the $x$--$y$ plane. ($a$,$d$) cuboidal canopy elements from case P1; T-shaped canopy elements from ($b$,$e$) case T1 and ($c$,$f$) case TP1. The white lines mark the positions of the canopy elements. The contours are scaled using the global friction velocity, $u_\tau$, of each case.}
	    	\label{fig:canopy_element_geom}	
 \end{figure}

{We have considered flows over canopies with both permeable and impermeable elements. The geometry of the impermeable canopy elements} is resolved using an immersed-boundary method adapted from \citet{Garcia-Mayoral2011}. The simulation parameters for the different cases studied here are summarised in table~\ref{tab:DNS_param}. Case S is an open channel with a smooth-wall floor. The canopy-resolving simulations include two canopy geometries, as portrayed in figure~\ref{fig:canopy_element_geom}, with varying element spacings. The first geometry, denoted by the letter `P', consists of collocated prismatic-posts with a square top-view cross-section with sides $\ell_x^+ = \ell_z^+ \approx 20$, and height $\ell_s^+ \approx 110$. {The spacing between the canopy elements in the wall-parallel directions for cases PD, P0, P1 and P2 are $L_x^+ = L_z^+ \approx 50, 100, \ 200$ and $400$, respectively. The canopy of case PD has a frontal area density of $\lambda_f \approx 0.88$, which would place it in the dense regime \citep{Nepf2012}. We use this simulation to contrast sparse and dense canopy dynamics.} The second geometry, denoted by the letter `T', consists of frontally-extruded T-shaped canopies, as portrayed in figures~\ref{fig:canopy_element_geom}($b$--$c$), {in a collocated arrangement}. {We consider two element spacings for the T-shaped canopies, with cases T1 and T2 having $L_{x}^+ = L_{z}^+ \approx 200$ and $400$, respectively}. The head of these canopy elements has dimensions $\ell_x^+ = \ell_z^+ \approx 40$ in the wall-parallel directions. The base of the canopy elements has $\ell_x^+ \approx 40$ and $\ell_z^+ \approx 20$. The base and the head are $\ell_s^+ \approx 80$ and $\ell_h^+ \approx 30$ tall, respectively. {We also study how canopies with permeable and impermeable canopy elements affect the surrounding flow. The permeable canopy elements of case} TP1 have the same geometry and layout as T1, but the elements are represented by a drag force only applied within them, rather than by immersed boundaries. This method allows some flow to permeate into the canopy elements, as shown in figures~\ref{fig:canopy_element_geom}($c$,$f$), and has been observed to be a suitable model for certain natural canopies \citep{Yan2017,Yue2007}. The drag force in case TP1, applied only at the grid points that are within the canopy elements, is of the form $C_{dc} {u_i} |{u_i}|$, similar to \citet{Yue2007}, \citet{Bailey2013} and \citet{Yan2017}, where $C_{dc}$ is a drag coefficient and ${u_i}$ is the instantaneous local velocity in every $i$ direction. $C_{dc}$ is set such that further increasing its magnitude does not significantly increase the net drag force on the mean flow. This forcing provides a local body force opposing the flow inside the canopy elements, and thus results in a small velocity within the canopy elements. The net mean drag force for this canopy is similar to that of the impermeable canopy, T1, as noted in table~\ref{tab:DNS_param}, in spite of the different character of the canopy elements.

{In order to ascertain the effect of the Reynolds number on the results, two additional simulations, P2I$_{Re}$ and P2O$_{Re}$, are conducted at $\Rey_\tau \approx 1000$. The canopy of P2I$_{Re}$ matches the parameters of the canopy of P2 in inner units, that is, element widths $\ell_x^+ = \ell_z^+ \approx 20$, height $\ell_s^+ \approx 110$ and element spacings $L_x^+ = L_z^+ \approx 400$. The channel to canopy height ratio for P2I$_{Re}$ is $\delta/\ell_s \approx 10$, whereas for case P2 the ratio is $\delta/\ell_s \approx 5$. The simulation P2I$_{Re}$ is conducted to verify that the channel height used is large enough not to constrain the canopy-layer dynamics. The canopy of P2O$_{Re}$ matches the parameters of the canopy of P2 in outer units, that is, $\ell_x/\delta = \ell_z/\delta \approx 0.04$, height $\ell_s/\delta \approx 0.2$ and element spacings $L_x/\delta = L_z/\delta \approx 0.8$. This simulation is conducted to assess Reynolds-number effects for a fixed canopy geometry.}

The roughness densities of the canopies are given in table~\ref{tab:DNS_param}. All the canopies studied lie within the sparse to transitional regime empirically demarcated by \citet{Nepf2012}, {except that of case PD, which lies in the dense regime}. The spanwise spacings between the sparse canopy elements are $L_z^+ \gtrsim 100$, which is comparable to, or larger than, the width of near-wall streaks, $\lambda_z^+ \approx 100$ \citep{Kline1967}. This implies that the canopies should be sparse from the point of view of the near-wall turbulent fluctuations as well. 

\begin{figure}
	\centering
		\subfloat{%
%  		 \tikzsetnextfilename{Cd_Ts}
  		 \includegraphics[scale=1]{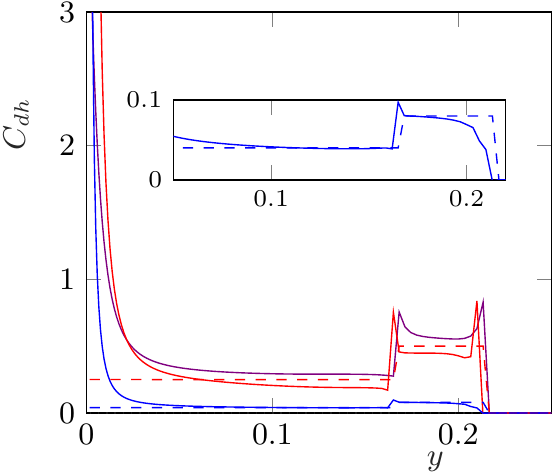}
  		}%
  		\caption{Drag coefficients, $C_{dh} = D/U^2$, obtained  from `T' shaped canopies. \protect\redline, case T1; \protect\violetline, case TP1; \protect\blueline, case T2; \protect\reddashed, cases T1-H/H0; and \protect\bluedashed, case T2-H0. The inset provides a magnified view of the drag coefficients for cases T2 and T2-H0.}
  		\label{fig:drag_coeff}
\end{figure}
\subsection{Drag force representations}
In order to explore the canopy-flow dynamics, we also conduct simulations where the canopy is replaced by some drag force, that does not resolve the geometry of the canopy elements. Sparse canopies consisting of bluff elements, such as those in the present work, are generally better characterised by a quadratic, {form} drag \citep{Coceal2008}, whereas for canopies {with slender, filamentous elements,} where viscous effects can dominate, a drag force proportional to the velocity would be more appropriate \citep{Tanino2008,Sharma2019dense}. {For complex natural canopies with foliage, which can have a range of element scales, both form and viscous drags can be important \citep{Finnigan2000}}. {For the canopies studied here, we find that the drag is essentially quadratic, and thus replace} the canopy elements by a quadratic drag force. {Note, however, that whether the drag was linear, quadratic, or else is of no consequence to the conclusions that we derive.}

The drag coefficient, $C_{dh}$, is obtained by approximating the canopy drag force obtained from the canopy-resolving simulations, $D$, to a form $D \approx C_{dh} U |U|$, where $U$ is the mean streamwise velocity. The drag coefficients obtained from cases T1, TP1 and T2 are portrayed in figure~\ref{fig:drag_coeff}. This quadratic form provides a reasonable approximation of the drag force for $y^+ \gtrsim 20$, once viscous effects are small. This is consistent with observations made in previous studies \citep{Coceal2006,Coceal2008,Bohm2013}. 

For the simulations labelled with the suffix `-H', the presence of the canopy is replaced by a force $C_{dh} {u_i} |{u_i}|$ applied homogeneously below the canopy tips. This is the conventional homogeneous-drag model. It also requires the prescription of drag coefficients in the spanwise and wall-normal directions. We estimate these by rescaling the streamwise drag coefficient based on the relative change in the `blockage ratio'  of the canopy elements in the different directions \citep{Luhar2008}, in the spirit of the method proposed by \citet{Luhar2013}. The blockage ratio in each direction is proportional to the frontal area of the canopy elements in that direction. In the wall-normal and spanwise directions this would be the top-view and the side-view areas respectively. For the wall-normal drag, this assumption is particularly coarse, but \citet{Busse2012} have shown that the flow is relatively insensitive to moderate changes in the wall-normal drag coefficient. In the simulations labelled with the suffix `-H0', a forcing $C_{dh}  {U}  |{U}| $ is applied in the region below the canopy tips, where $U(y)$ is the mean velocity profile. The drag is only applied to the mean-streamwise velocity, and has no fluctuating component. While the drag force in cases labelled `-H' varies along any given wall-parallel plane depending on the local velocity, in cases labelled `-H0' the drag force is homogeneous along any given wall-parallel plane, as it depends only on the mean velocity and the drag coefficient at that height. {Note that as the aforementioned drag models do not resolve the shape of the canopy elements, they also cannot capture the element-induced flow. In order to capture a part of the effect of the element-induced flow,} the simulation labelled with the suffix `-L' applies a drag $C_{dh} {U} | {U} | $, as in cases H0, but distributed in a reduced-order representation of the canopy elements. This representation consists of a 24-mode, $x$-$z$ Fourier-truncation of the canopy geometry. 

\subsection{Effect of Reynolds number}\label{subsec:re_effect}

\begin{figure}
	\centering
	\subfloat{%
%  		 \tikzsetnextfilename{urms_P_pg_Re}
%  		\input{img_Re/u_c_p_g.tex}
  		\includegraphics[scale=1]{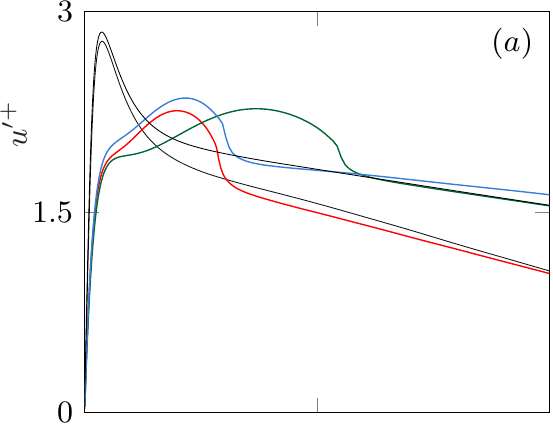}
  		}%
         \hspace{0.5cm}\subfloat{%
%  		 \tikzsetnextfilename{urms_P_gg_Re}
%  		\input{img_Re/u_c_g_g.tex}
  		\includegraphics[scale=1]{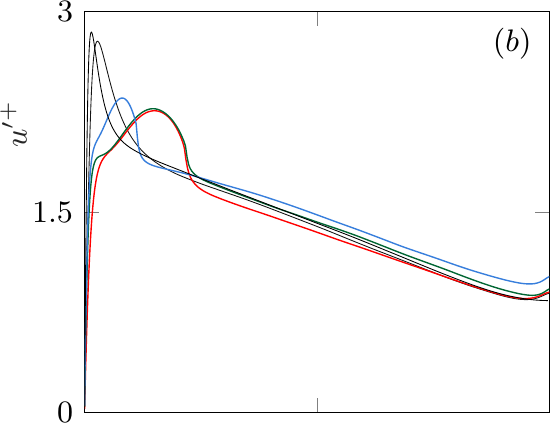}
  		}%
  		
       \vspace{0mm}\subfloat{%
%  		 \tikzsetnextfilename{vrms_P_pg_Re}
%  		\input{img_Re/v_c_p_g.tex}
  		\includegraphics[scale=1]{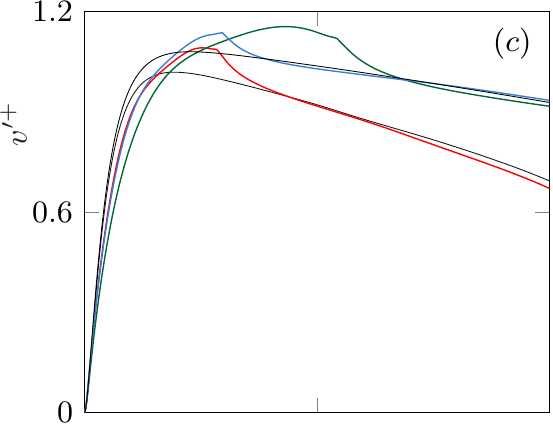}
  		}% 
       \hspace{0.5cm}\subfloat{%
%  		 \tikzsetnextfilename{vrms_P_gg_Re}
%  		\input{img_Re/v_c_g_g.tex}
  		\includegraphics[scale=1]{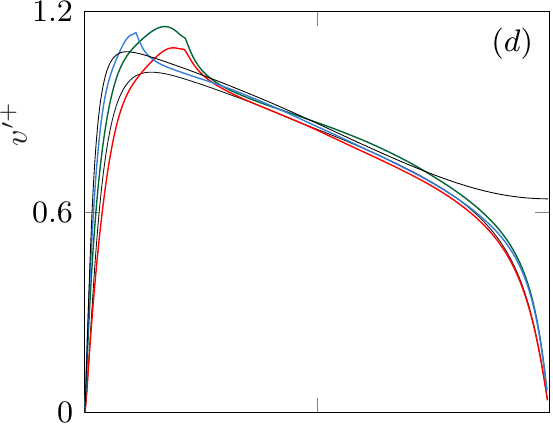}
  		}%

  		\vspace{3mm}
  		 \hspace{0.04cm}\subfloat{%
% 		\tikzsetnextfilename{wrms_P_pg_Re}
% 	   \input{img_Re/w_c_p_g.tex}
 	   \includegraphics[scale=1]{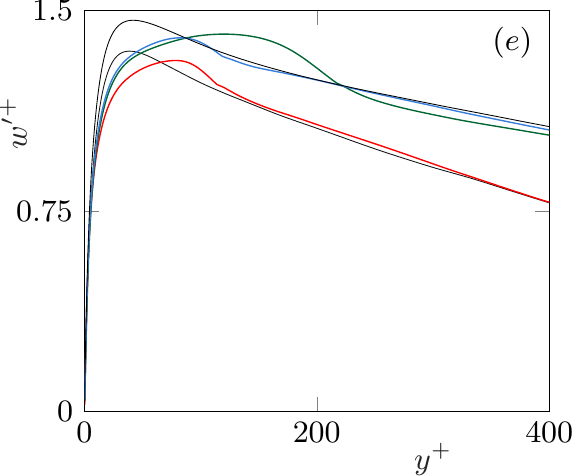}
 		}%	
  		\hspace{0.28cm}\subfloat{%
% 		\tikzsetnextfilename{wrms_P_gg_Re}
% 		\input{img_Re/w_c_g_g.tex}
 		\includegraphics[scale=1]{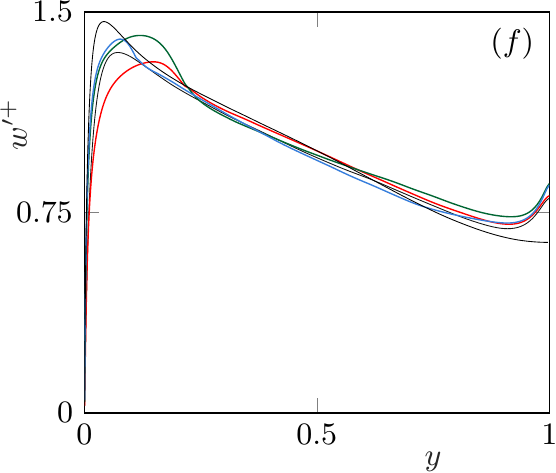}
 		}%

  	 \caption{{Rms velocity fluctuations scaled with the global friction velocity, $u_\tau$, from \protect\redline, case P2; \protect\blueReline, case P2I$_{Re}$; and \protect\darkgreenline, case P2O$_{Re}$. In the left column, the wall-normal coordinate is scaled in friction units, and in the right column, in outer units. The black lines represent the smooth-wall simulations at $\Rey_\tau \approx 520$ and $1000$. The smooth-wall data at $Re_\tau \approx 1000$ is taken from \citet{Lee2015}.}}%
	    	\label{fig:stats_resolved_g_Re}	
\end{figure} 

\begin{figure}
	\centering
	    \subfloat{%
% 		\tikzsetnextfilename{uvrms_P_pg_Re}
% 		\input{img_Re/Tau_c_p_g.tex}
 		\includegraphics[scale=1]{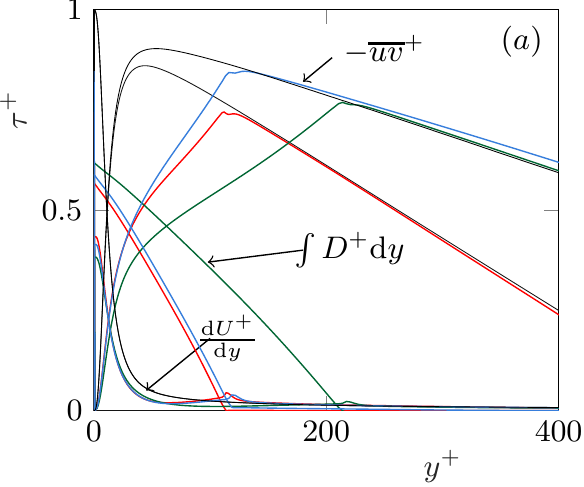}
 		}%
  		\hspace{3mm}\subfloat{%
% 		\tikzsetnextfilename{uvrms_P_gg_Re}
% 		\input{img_Re/Tau_c_g_g.tex}
 		\includegraphics[scale=1]{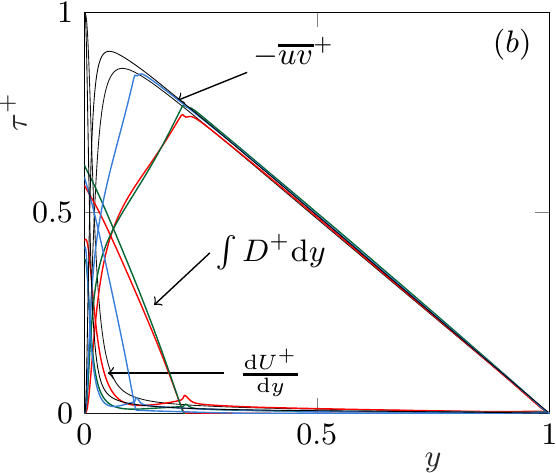}
 		}%
  	 \caption{{Viscous and Reynolds shear stresses scaled with the global friction velocity, $u_\tau$, from \protect\redline, case P2; \protect\blueReline, case P2I$_{Re}$; and \protect\darkgreenline, case P2O$_{Re}$. In the left column, the wall-normal coordinate is scaled in friction units, and in the right column, in outer units. The black lines represent the smooth-wall simulations at $\Rey_\tau \approx 520$ and $1000$. The smooth-wall data at $Re_\tau \approx 1000$ is taken from \citet{Lee2015}.}}%
  	 	\label{fig:stress_resolved_g_Re}	
\end{figure}

\begin{figure}
	\centering
	  	 \subfloat{%
%  		 \tikzsetnextfilename{Cdp_vs_yp_Re}
%  		 \input{img_Re/Cdp_vs_yp.tex}
  		 \includegraphics[scale=1]{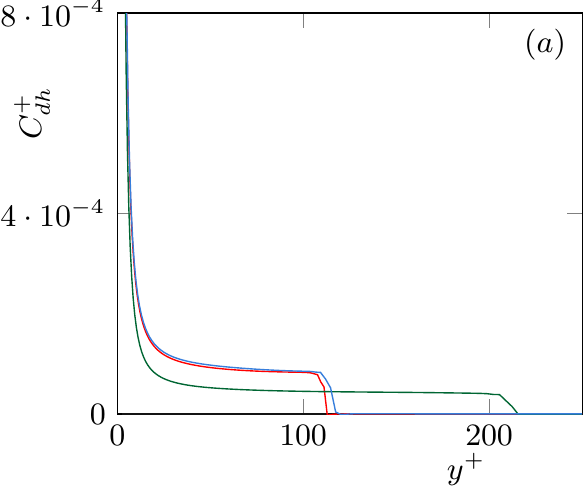}
  		}%
		\hspace{3mm}\subfloat{%
%  		 \tikzsetnextfilename{Cd_vs_y_Re}
%  		\input{img_Re/Cd_vs_y.tex}
  		\includegraphics[scale=1]{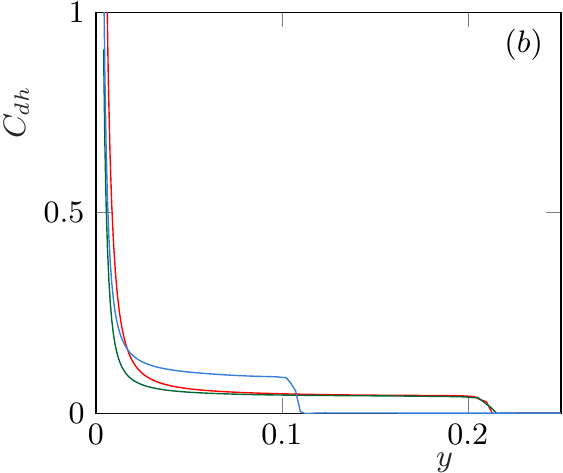}
  		}%
  		\caption{{Drag coefficients, $C_{dh} = D/U^2$ for \protect\redline, case P2; \protect\blueReline, case P2I$_{Re}$; and \protect\darkgreenline, case P2O$_{Re}$.}}
  		\label{fig:drag_coeff_Re}
\end{figure}

{In order to analyse the effect of the Reynolds number on the DNS results presented in the subsequent sections, we first compare the results of case P2 to those of cases P2I$_{Re}$ and P2O$_{Re}$. The simulations P2 and P2I$_{Re}$ have the same canopy parameters in friction units and $\Rey_\tau \approx 520$ and $1000$, respectively. The velocity fluctuations and the Reynolds shear stresses for these simulations essentially collapse up to a height $y^+ \approx 20$, as shown in figures~\ref{fig:stats_resolved_g_Re}($a,c,e$) and figure~\ref{fig:stress_resolved_g_Re}($a$). At heights larger than $y^+ \gtrsim 20$, the magnitudes of velocity fluctuations and the Reynolds shear stresses are larger for case P2I$_{Re}$ than for case P2. This behaviour is consistent with that observed for smooth-wall flows at the corresponding Reynolds numbers \citep{Sillero2013}, which suggests that the differences in the velocity fluctuations observed at heights $y^+ \gtrsim 20$ do not result from the presence of the canopy. At heights of $y^+ \gtrsim 200$, the velocity fluctuations of the canopy simulations collapse with those of their respective smooth-wall simulations, which indicates the recovery of outer-layer similarity. In addition, the effective canopy drag coefficients, $C_{dh}$, for the canopies of cases P2 and P2I$_{Re}$ also collapse in friction units, as shown in figure~\ref{fig:drag_coeff_Re}($a$). These results show that the domain height used in case P2 is sufficiently large and does not constrain the flow within the canopy-layer. Increasing the domain height further simply results in a larger region above the canopy-layer exhibiting outer-layer similarity. This is consistent with the study of flows over cube canopies by \citet{Coceal2006}, who also noted that increasing the height of the domain beyond $\delta/\ell_s \approx 4$ did not have a significant effect on the flow within the canopy-layer.}

{We now compare the results from cases P2 and P2O$_{Re}$, which have the same canopy parameters when scaled in outer units, but different friction Reynolds numbers. The canopy heights for both of these cases is $\ell_s/\delta \approx 0.2$, and in both cases the elements extend well into the logarithmic, self-similar region of the flow. Close to the wall, $y^+ \lesssim 20$, the velocity fluctuations and Reynolds shear stresses for these cases collapse when scaled in friction units. At larger heights, $y^+ > 20$, the streamwise velocity fluctuations and the Reynolds shear stresses essentially collapse when the wall-normal coordinate is scaled in outer units, as shown in figures~\ref{fig:stats_resolved_g_Re}($b$) and \ref{fig:stress_resolved_g_Re}($b$). The cross velocity fluctuations in the region between $y^+ \approx 20$ and $y/\delta \approx 0.2$ are larger for case P2O$_{Re}$ compared to case P2, as illustrated in figures~\ref{fig:stats_resolved_g_Re}($d$) and ($f$). This increase, however, is also observed in the cross velocity fluctuations of the corresponding smooth-wall simulations. Beyond a height of $y/\delta \approx 0.25$, we observe that the velocity fluctuation and Reynolds shear stress profiles for cases P2, P2O$_{Re}$, P2I$_{Re}$ and the smooth-wall simulations coincide. Furthermore, the effective canopy drag coefficient profiles for cases P2 and P2O$_{Re}$, portrayed in figure~\ref{fig:drag_coeff_Re}($b$), also collapse when scaled in outer units, consistent with the observations of \citet{Cheng2002} for cube canopies. Therefore, the drag coefficient of the canopy is essentially independent of the Reynolds number, implying that the canopy is in the fully-rough regime \citep{Nikuradse1933}. These results therefore suggest that the conclusions drawn in the following sections from simulations at $\Rey_\tau \approx 520$ should also be relevant for higher Reynolds number flows. Note also that the quotient $D/U^2$ becomes close to constant for $y^+ \gtrsim 25$, indicating that the total canopy drag, $D$, is essentially quadratic above this height.}

\section{Canopy-resolving simulations}\label{sec:resolved_canopy}

In this section, we present and discuss the scaling of turbulent fluctuations in sparse canopies, and compare them with those over a smooth wall. Over a smooth wall, the balance of stresses within the channel can be obtained by averaging the momentum equations in the wall-parallel directions and time and integrating in $y$, which yields
\begin{eqnarray}
 \frac{\mathrm{d} P}{\mathrm{d} x} y + \tau_w & = & -\overline{u v} + \nu\frac{\mathrm{d} U}{\mathrm{d} y},
  \label{eq:stress_balance_smooth}
 \end{eqnarray}
where $\tau_w$ is the wall shear stress, $\mathrm{d} P/\mathrm{d} x$ is the mean streamwise pressure gradient, $-\overline{u v}$ is the Reynolds shear stress, $U$ is the mean streamwise velocity, and $\nu$ is the kinematic viscosity. Particularising \eqref{eq:stress_balance_smooth} at $y = \delta$, we obtain the expression for the wall shear stress and the friction velocity, $u_\tau$,
\begin{eqnarray} \label{eq:utau_smooth}
 u_\tau^2  =  \tau_w  = -\delta \frac{\mathrm{d} P}{\mathrm{d} x}.
\end{eqnarray}
In the presence of a canopy, the stress balance also includes the drag exerted by the canopy elements,
\begin{eqnarray}\label{eq:stress_balance_canopy}
 \frac{\mathrm{d} P}{\mathrm{d} x} y + \tau_w  & = & -\overline{u v} + \nu\frac{\mathrm{d} U}{\mathrm{d} y} - \int_{0}^{y} D \ \mathrm{d} y,
\end{eqnarray}
where the canopy drag averaged in $x$, $z$ and time, $D$, is zero in {the region above the canopy tips}, $y > h$. {This stress balance is typically used in canopy literature to calculate the canopy drag stress \citep{Dunn1996, Ghisalberti2004}.} Equation~\eqref{eq:stress_balance_canopy} can be rewritten as
\begin{eqnarray}\label{eq:stress_balance_canopy2}
 \frac{\mathrm{d} P}{\mathrm{d} x} y + \tau_w + \int_{0}^{h} D \ \mathrm{d} y & = & -\overline{u v} + \nu\frac{\mathrm{d} U}{\mathrm{d} y} + \int_{y}^{h} D \ \mathrm{d} y,
 \end{eqnarray}
so that the net drag, $\tau_w + \int_{0}^{h} D \ \mathrm{d} y$, is on the left-hand side, as in \eqref{eq:stress_balance_smooth}. 
From this net drag, a `global' friction velocity can be defined, 
\begin{eqnarray} \label{eq:global_utau}
 u_\tau^2 = \tau_w + \int_{0}^{h} D \ \mathrm{d} y & = & -\delta \frac{\mathrm{d} P}{\mathrm{d} x}.
 \end{eqnarray} 
 \begin{figure}
        \centering
         \includegraphics[scale=0.325,trim={0mm 0mm 0mm 0mm},clip]{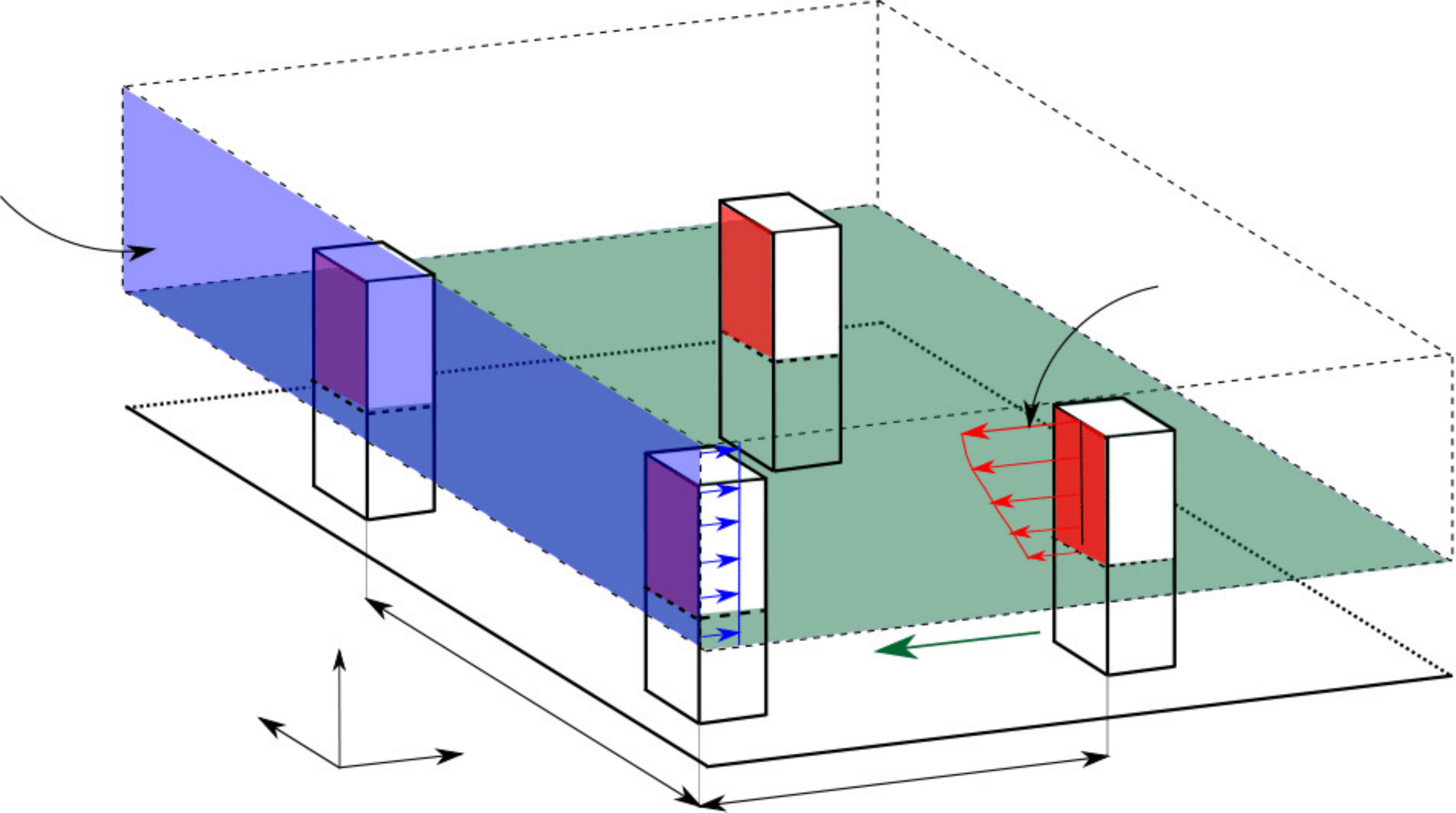}%
        \mylab{-11.5cm}{4.8cm}{\textcolor{blue}{$-\frac{\mathrm{d}P}{\mathrm{d}x}$}}%
        \mylab{-2.2cm}{3.8cm}{\textcolor{red}{$D(y)$}}%
        \mylab{-4.8cm}{0.75cm}{\rotatebox{5}{\textcolor{darkgreen}{$-\overline{uv} + \nu\frac{\mathrm{d}U}{\mathrm{d}y}$}}}%
          \mylab{-7.6cm}{0.1cm}{$x$}% 
          \mylab{-8.2cm}{1cm}{$y$}% 
          \mylab{-9.1cm}{0.3cm}{$z$}% 
           \mylab{-4cm}{0cm}{$s$}% 
            \mylab{-7.1cm}{0.6cm}{$s$}% 
          \caption{{Schematic representation of the stress balance in a channel with canopy elements.}}%
            \label{fig:stress_balance_schematic}
\end{figure} 
 \begin{figure}
		\centering
         \subfloat{%
%  		 \tikzsetnextfilename{stress_balance}
%  			\input{img_16x8_posts/stress_balance_g.tex}
  			\includegraphics[scale=1]{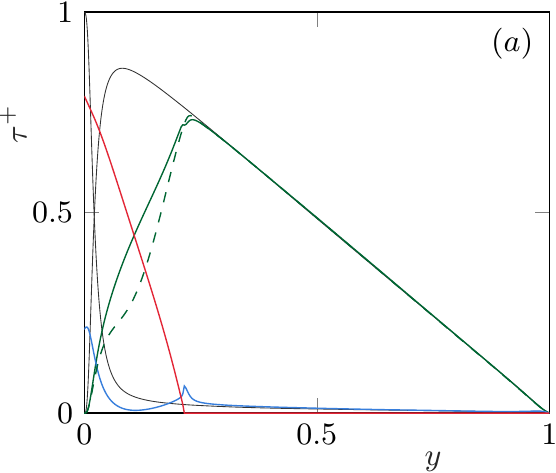}
  		}%
  		\hspace{3mm}\subfloat{%
%  		 \tikzsetnextfilename{tau_resc}
%  			\input{img_16x8_posts/uv_dudy_ul_yg.tex}
  			\includegraphics[scale=1]{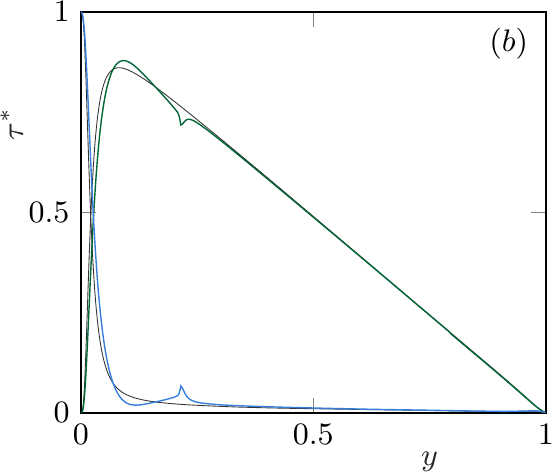}
  		}%
  	\caption{{Stress profiles within the channel for case P1. (\textit{a}) \protect\darkgreenline, full Reynolds shear stress; \protect\darkgreendashed, background turbulent Reynolds shear stress; \protect\redline, drag stress; and \protect\blueReline, viscous stress; scaled with $u_{\tau}$. (\textit{b}) \protect\darkgreenline, background turbulent Reynolds shear stress; and \protect\blueReline, viscous stress; scaled with $u^*$. The black lines represent the the smooth-wall case, S.}}
	    	\label{fig:stress_profiles}	
\end{figure}  
This is the equivalent of the smooth-wall $u_\tau$ of equation~\eqref{eq:utau_smooth} for canopy flows.

{While in smooth-wall flows the total stress is the sum of the viscous and Reynolds shear stresses alone and is linear in $y$, in canopy flows that linear sum has an additional contribution from the canopy drag as evidenced by the right-hand-side of equation~\eqref{eq:stress_balance_canopy2}. This equation also portrays that, at any given height $y$, the sum of the streamwise shear stresses, $-\overline{uv} + \nu\mathrm{d}U/\mathrm{d}y$, and the drag from the canopy above that height, $\int_{y}^{h} D \ \mathrm{d} y$, are balanced by the force exerted by the pressure gradient above. This can also be obtained from an integral balance of forces between heights $y$ and $\delta$, and is illustrated by the sketch in figure~\ref{fig:stress_balance_schematic}. Outside the canopy, the drag term is zero, and the magnitude of the viscous and Reynolds shear stresses is similar to that over smooth walls. Within the canopy, however, the canopy drag can dominate, and the viscous and Reynolds shear stresses are smaller than over smooth walls, as shown in figure~\ref{fig:stress_profiles}(\textit{a}).}

\begin{figure}
	\centering
         \subfloat{%
%  		 \tikzsetnextfilename{urms_T_g_16x8}
%  		\input{img_16x8/u_c_g.tex}
  		\includegraphics[scale=1]{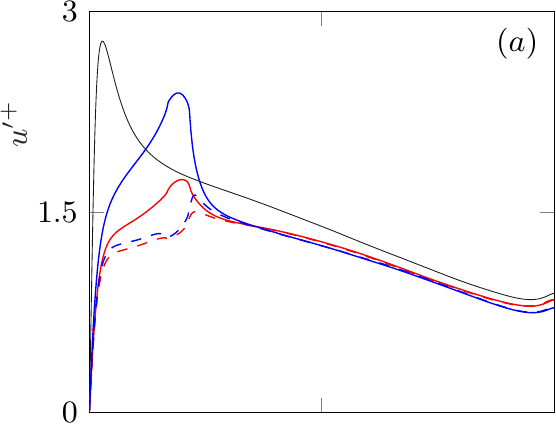}
  		}%  		
         \hspace{3mm}\subfloat{%
%  		 \tikzsetnextfilename{vrms_T_g_16x8}
%  		\input{img_16x8/v_c_g.tex}
  		\includegraphics[scale=1]{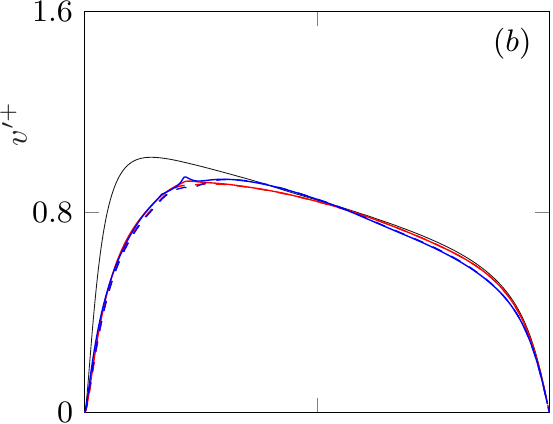}
  		}%
  		
       \vspace{3mm}\hspace{1mm}\subfloat{%
% 		\tikzsetnextfilename{wrms_T_g_16x8}
% 		\input{img_16x8/w_c_g.tex}
 		\includegraphics[scale=1]{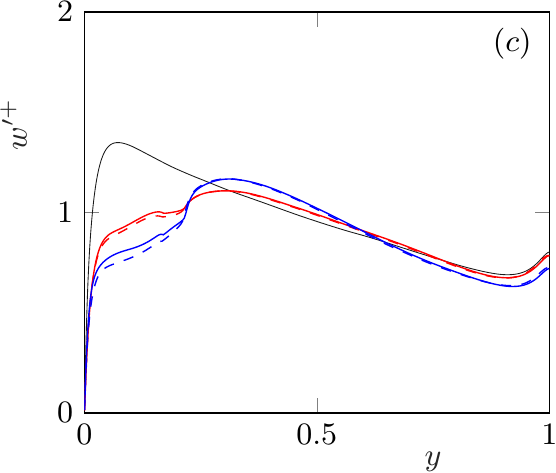}
 		}%		
        \hspace{2.5mm}\subfloat{%
% 		\tikzsetnextfilename{uvrms_T_g_16x8}
% 		\input{img_16x8/Tau_c.tex}
 		\includegraphics[scale=1]{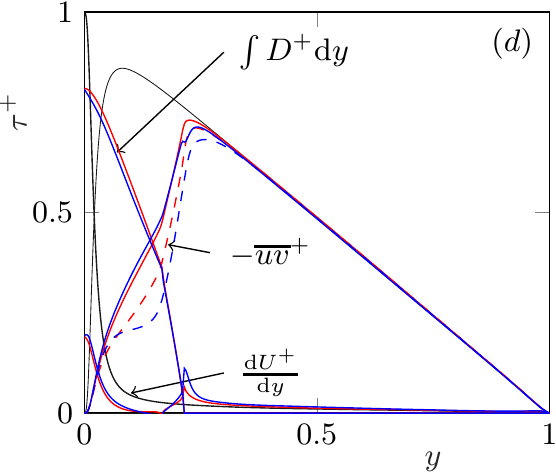}
 		}%
 \caption{Rms velocity fluctuations and shear stresses scaled with the global friction velocity, $u_\tau$, for \protect\redline, case TP1; and \protect\blueline, case T1. Solid lines represent the full velocity fluctuations and dashed lines represent the background-turbulence fluctuations. The black lines represent the smooth-wall case, S, for reference.}%
	    	\label{fig:stats_resolved_g_16x8}%
\end{figure}

\begin{figure}
	\centering
         \subfloat{%
%  		 \tikzsetnextfilename{urms_P_g}
%  		\input{images_posts/u_P_g.tex}
  		\includegraphics[scale=1]{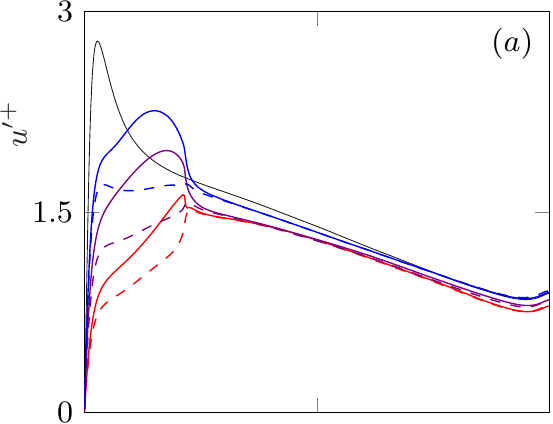}
  		}%
  		\hspace{3mm}\subfloat{%
%  		 \tikzsetnextfilename{urms_P_l}
%  		\input{images_PT_local/u_P_l.tex}
  		\includegraphics[scale=1]{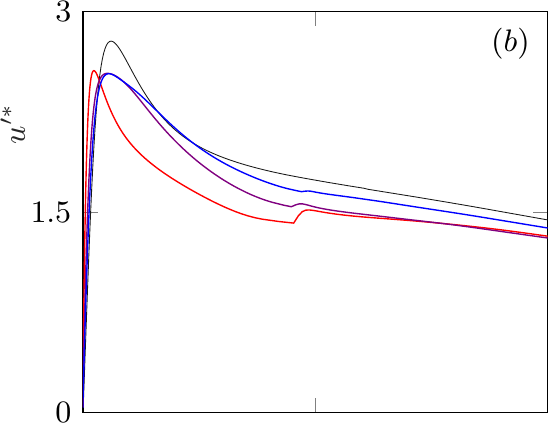}
  		}%
  		
       \vspace{0mm}\subfloat{%
%  		 \tikzsetnextfilename{vrms_P_g}
%  		\input{images_posts/v_P_g.tex}
  		\includegraphics[scale=1]{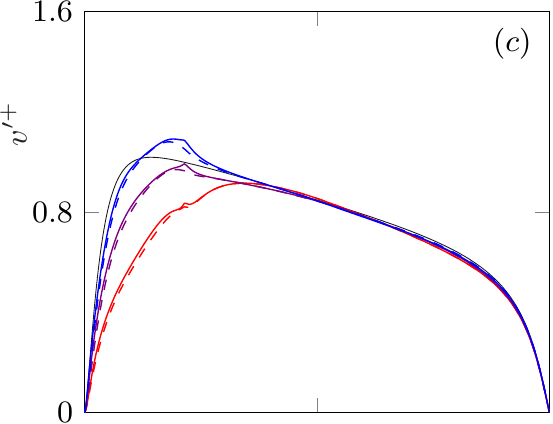}
  		}%
  		\hspace{3mm}\subfloat{%
%  		 \tikzsetnextfilename{vrms_P_l}
%  		\input{images_PT_local/v_P_l.tex}
  		\includegraphics[scale=1]{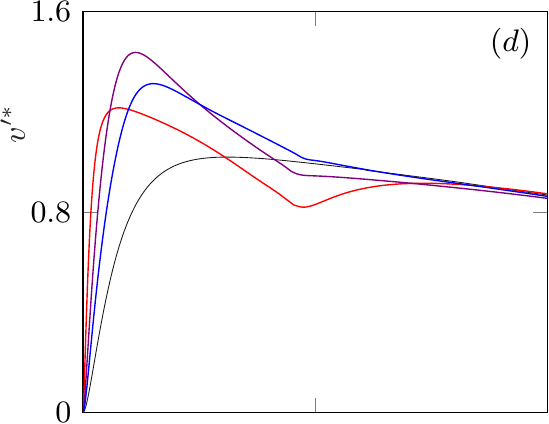}
  		}% 
  		
  		\vspace{3mm}\hspace{2.3mm}\subfloat{%
% 		\tikzsetnextfilename{wrms_P_g}
% 		\input{images_posts/w_P_g.tex}
 		\includegraphics[scale=1]{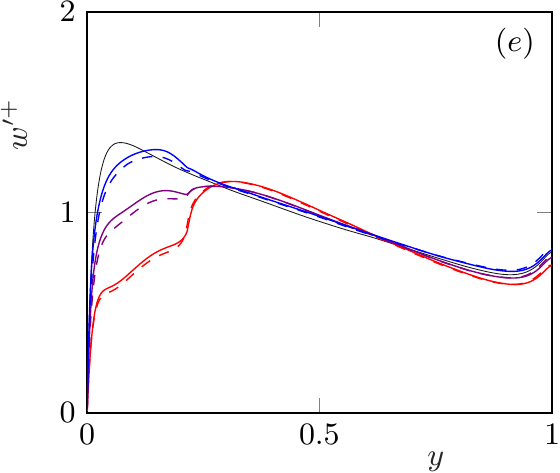}
 		}%
       \hspace{1.5mm} \subfloat{%
% 		\tikzsetnextfilename{wrms_P_l}
% 	   \input{images_PT_local/w_P_l.tex}
 	   \includegraphics[scale=1]{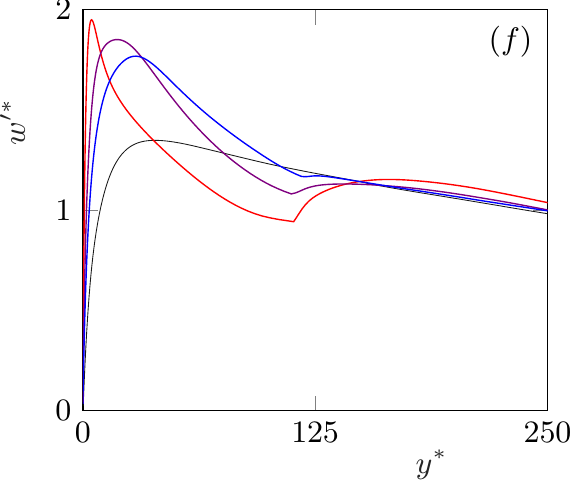}
 		}%	
  	 \caption{{Rms velocity fluctuations scaled with the global friction velocity, $u_\tau$,  in the left column and with the local friction velocity, $u^*$, in the right column. The lines represent \protect\redline, case P0; \protect\violetline, case P1; and \protect\blueline, case P2. In the left column, solid lines represent the full velocity fluctuations and dashed lines represent the background-turbulence fluctuations. In the right column, only the background-turbulent fluctuations are portrayed. The black lines represent the smooth-wall case, S, for reference.}}%
	    	\label{fig:stats_resolved_gl_Posts}	
\end{figure}

 \begin{figure}
		\centering
         \subfloat{%
%  		 \tikzsetnextfilename{y_vs_yl}
%  			\input{images_posts/y_vs_yl_F.tex}
  			\includegraphics[scale=1]{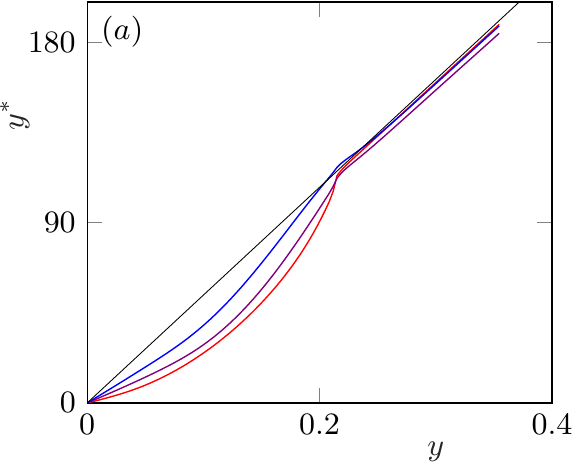}
  		}%
         \hspace{3mm}\subfloat{%
%  		 \tikzsetnextfilename{y_vs_ul}
%  			\input{images_posts/y_vs_ul_F.tex}
  			\includegraphics[scale=1]{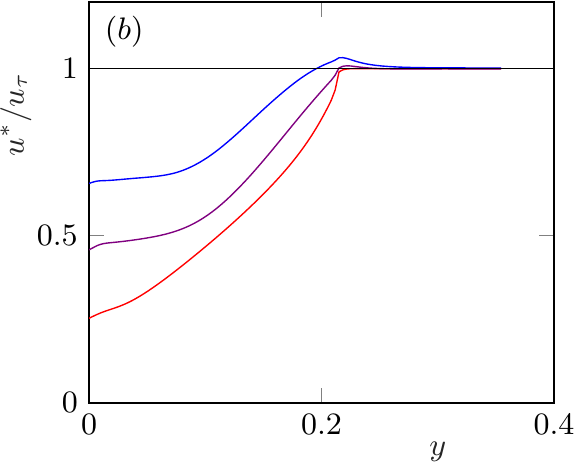}
  		}%
  	\caption{Variation of ($a$) $y^*$, and ($b$) $u^*$ with height for cuboidal post canopies, with sparsity increasing from red to blue. \protect\redline, denser case P0; \protect\violetline, intermediate case P1; \protect\blueline, sparser case P2; and \protect\blackthin, smooth-wall S. }
	    	\label{fig:local_scaling}	
\end{figure}

\begin{figure}
	\centering
  		\subfloat{%
% 		\tikzsetnextfilename{uvrms_P_g}
% 		\input{images_posts/Tau_P.tex}
 		\includegraphics[scale=1]{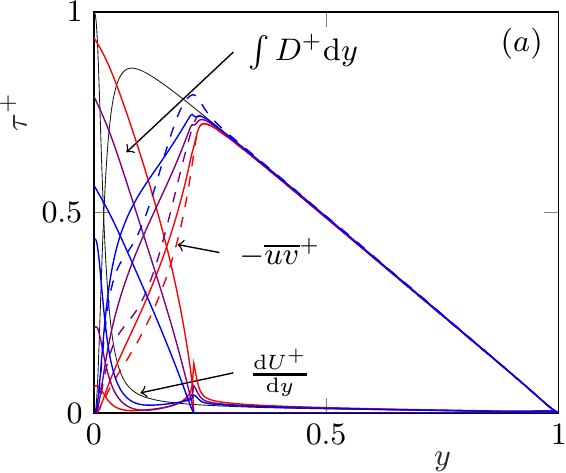}
 		}%
       \hspace{3mm}\subfloat{%
% 		\tikzsetnextfilename{uvrms_P_l}
% 		\input{images_PT_Local/Re_stress_P_l.tex}
 		\includegraphics[scale=1]{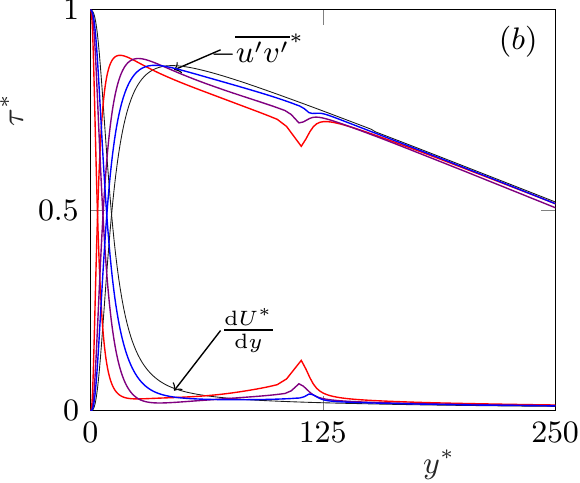}
 		}%
  	 \caption{Shear stresses scaled using ($a$) the global friction velocity, $u_\tau$ and ($b$) the local friction velocity, $u^*$, for \protect\redline, case P0; \protect\violetline, case P1; and \protect\blueline, case P2. The solid and dashed lines in ($a$) represent the full and background turbulent Reynolds shear stresses, respectively. In ($b$) only the background-turbulent Reynolds shear stresses are portrayed. The black lines represent the smooth-wall case, S.}%
  	 	\label{fig:stress_resolved_gl_Posts}	
\end{figure} 

As can be observed in figure~\ref{fig:channel_schematic}, the presence of the canopy elements induces a coherent flow. Several studies have shown that the flow around the canopy elements and the flow far away from them have significantly different characteristics, and consequently they are typically studied separately \citep{Finnigan2000,Bohm2013,Bailey2013}. A commonly used technique to separate the element-induced flow from the background turbulence is through triple decomposition \citep{Reynolds1972}
\begin{equation}
\boldsymbol{u} = \boldsymbol{U} + \boldsymbol{\widetilde{u}} + \boldsymbol{u'},
\end{equation}
where $\boldsymbol{u}$ is the full velocity. The mean velocity, $\boldsymbol{U}$, is obtained by averaging the flow in time and space. The element-induced velocity, also referred to as the dispersive flow, $\boldsymbol{\widetilde{u}}$, is obtained by {ensemble-}averaging the flow in time alone. We refer to the remaining part of the flow, $\boldsymbol{u'}$, as the incoherent, background-turbulence velocity. {Similarly, we refer to the Reynolds shear stress calculated using the full velocity, {$\overline{uv}$}, as the `full' Reynolds shear stress and that calculated using the incoherent, background-turbulence velocity, {$\overline{u'v'}$}, as the background-turbulence Reynolds shear stress. The difference between these two stresses gives a measure of the element-induced, dispersive stress,} {$\overline{\tilde{u}\tilde{v}} = \overline{uv} - \overline{u'v'}$. Note that this is slightly different from the commonly used notation, where the dispersive and background-turbulence Reynolds shear stresses are treated distinctly and the `full' Reynolds shear stress is not labelled \citep[e.g.,][]{Coceal2006}.} Essentially identical to triple decomposition, `double averaging' \citep{Raupach1982} can also be used to separate the element-induced and background-turbulence flows \citep{Finnigan2000,Nepf2012,Bai2015,Giometto2016,Yan2017}. 

The intensity of the element-induced flow can vary with the shape \citep{Balachandar1997,Taylor2011}, permeability \citep{Yu2010,Ledda2018}, and distribution of the canopy elements. It is possible, however, for canopies to have different element-induced flows but similar background turbulence. To illustrate this, we compare two canopy simulations, T1 and TP1, which have similar canopy layouts and roughly the same net drag, as shown in figure~\ref{fig:stats_resolved_g_16x8}($d$). The difference between the two cases is that in T1 the canopy elements are impermeable, whereas in TP1 some flow penetrates into the elements. The rms fluctuations of the full and background-turbulence velocity components for these cases are shown in figure~\ref{fig:stats_resolved_g_16x8}. The magnitude of the full streamwise fluctuations within the canopy is significantly larger for T1 than for TP1. This increase, however, can be attributed essentially to the stronger element-induced fluctuations generated by the impermeable canopy elements. This is evidenced by the fact that the background-turbulence streamwise fluctuations for both cases essentially collapse, as shown in figure~\ref{fig:stats_resolved_g_16x8}($a$). The cross-flow fluctuations and Reynolds shear stress profiles for both these canopies are also similar. The impermeable canopy, however, has a slightly larger damping effect on the spanwise fluctuations.

The fluctuating velocities portrayed in figures~\ref{fig:stats_resolved_gl_Posts}($a,c,e$) are scaled using the `global'  friction velocity defined by equation~\eqref{eq:global_utau}, which includes the full contribution of the canopy drag. \citet{Tuerke2013} studied smooth-wall flows with artificially forced mean profiles, and observed that the turbulent fluctuations in such flows scaled with the local sum of the viscous and Reynolds shear stresses, or the local stress $\tau_f$, at each height. This was the case even when $\tau_f$ was not linear with $y$ due to the artificial forcing. {This idea has been expanded on by \citet{Lozano2019}, who proposed that the energy containing turbulent scales in the logarithmic layer in smooth-wall flows also scale with local velocity and length scales at each height, irrespective of the location of the wall.} \citet{Tuerke2013} defined a `local' friction velocity, $u^*$, by linearly extrapolating the local stress at each height to the wall,
\begin{equation}
{u^*(y)}^2 = \frac{\delta}{\delta - y} \tau_f(y).
\end{equation}
\noindent Notice that, for a smooth unforced channel, $u^* = u_\tau$ at every height. Following \citet{Tuerke2013}, we define the sum of the viscous and background-turbulence Reynolds stresses as the `fluid' stress $\tau_f$. In the present work, we only discuss the scaling of the background-turbulence fluctuations. Hence, only the contribution of the background-turbulence Reynolds shear stresses to $\tau_f$ is considered. A similar concept was also proposed by \citet{Hogstrom1982} for flows over urban canopies. They scaled turbulence with a local friction velocity, defined as the square root of the magnitude of the local Reynolds shear stress, but had measurements only at heights where the contribution of the viscous stress to $\tau_f$ would be small. Using $u^*$, a local viscous lengthscale can also be defined, $\nu/u^*$,  and from it an effective viscous height, $y^* = yu^*/\nu$. Both $u^*$ and $y^*$ are portrayed, for the prismatic-post canopies, in figure~\ref{fig:local_scaling}. Near the canopy tips, where the element-induced drag is no longer present, the local friction velocity, $u^*$ becomes equal to the global $u_\tau$, and $y^*$ becomes equal to $y^+$. Making the canopy sparser reduces the canopy drag, and hence the difference between $u^*$ and $u_\tau$ within the canopy reduces with increasing canopy sparsity.

When scaled with $u_\tau$, as is done conventionally, the viscous and Reynolds shear stresses near the base of the canopy are highly damped compared to smooth walls. However, the balance of these stresses in $\tau_f$ remains close to that over smooth walls. This is illustrated in {figures~\ref{fig:stress_profiles}($b$) and \ref{fig:stress_resolved_gl_Posts}($b$), which portray} the terms in the stress balance within a channel with canopies scaled with $u_\tau$ and $u^*$. The similarity of the viscous and Reynolds shear stresses in the canopy-flow and smooth-wall cases suggests that the canopy acts on the background turbulence essentially through changing their local scale, rather than through a direct interaction of the canopy elements with the flow. To explore the scaling further, the background-turbulence rms fluctuations {for the prismatic post canopies} are portrayed scaled with $u^*$ in figures~\ref{fig:stats_resolved_gl_Posts}($b,d,f$). Scaling the fluctuations with the conventional $u_\tau$ shows a reduction of the fluctuations within the canopy compared to a smooth wall, as shown in figure~\ref{fig:stats_resolved_gl_Posts}($a,c,e$). With our proposed scaling with $u^*$, in contrast, the streamwise fluctuations appear similar to those in a smooth channel. The increase in spanwise and wall-normal fluctuations, shown in figures~\ref{fig:stats_resolved_gl_Posts}($b,d,f$), suggests however, that there is a relative increase in the intensity of the cross flow within the canopy compared to a smooth channel. {The velocity fluctuations and the shear stresses for the larger Reynolds number simulations and for the T-shaped canopies also exhibit similar behaviour, and are provided in appendix~\ref{appA} for reference.}

\begin{figure}
		\centering
         \vspace{0.5cm}\includegraphics[scale=1.0,trim={0mm 2.5mm 0mm 0mm},clip]{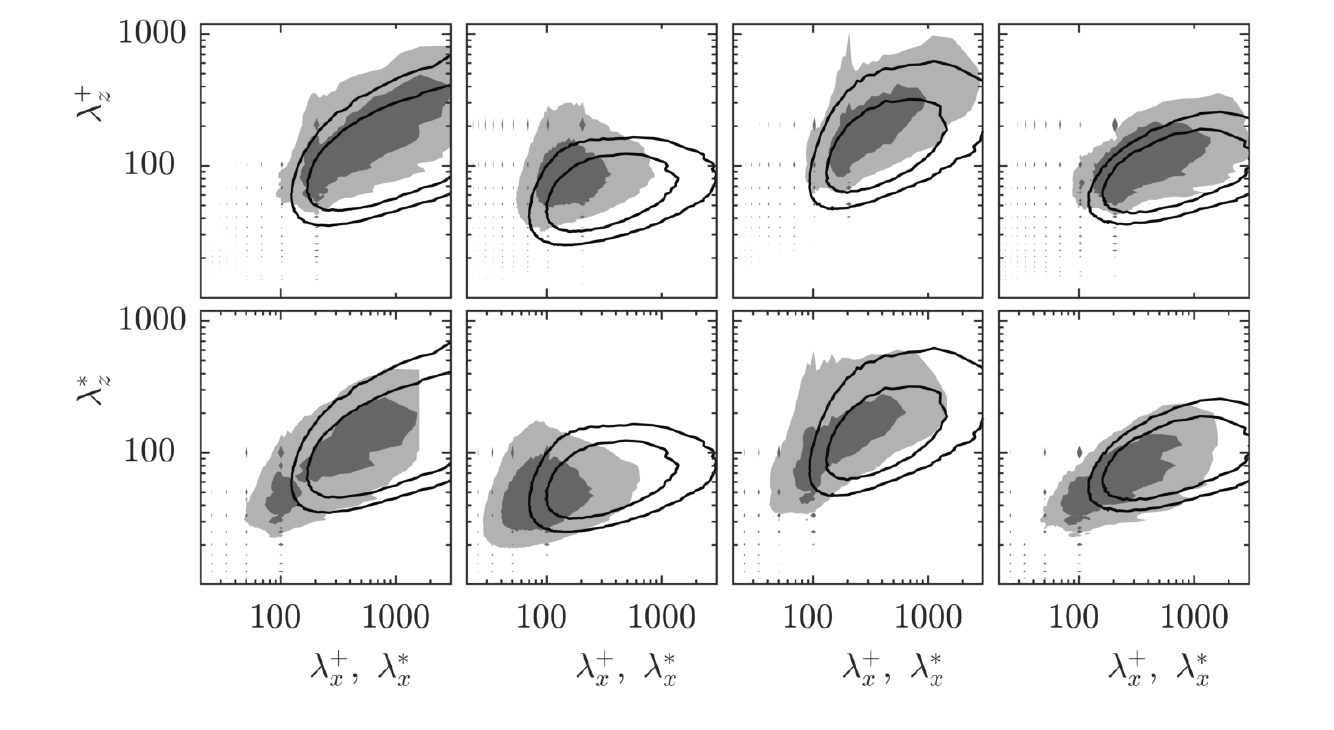}%
         \mylab{-11.45cm}{6.6cm}{(\textit{a})}%
  		\mylab{-8.7cm}{6.6cm}{(\textit{b})}%
  		\mylab{-6.0cm}{6.6cm}{(\textit{c})}%
  		\mylab{-3.3cm}{6.6cm}{(\textit{d})}%
         \mylab{-11.45cm}{3.70cm}{(\textit{e})}%
  		\mylab{-8.7cm}{3.70cm}{(\textit{f})}%
  		\mylab{-6.0cm}{3.70cm}{(\textit{g})}%
  		\mylab{-3.3cm}{3.70cm}{(\textit{h})}%
  		\mylab{-10.8cm}{7.0cm}{$k_x k_z E_{uu}$}%
   		 \mylab{-8.0cm}{7.0cm}{$k_x k_z E_{vv}$}%
   		 \mylab{-5.4cm}{7.0cm}{$k_x k_z E_{ww}$}% 	
   		 \mylab{-2.8cm}{7.0cm}{$k_x k_z E_{uv}$}% 	
		\caption{Pre-multiplied spectral energy densities, for case P1 (filled contours), and  for case S (line contours) normalised with their respective rms values at (\textit{a--d}) $y^+ = 15$ , and (\textit{e--h}) $y^* = 15$. The contours, from the left to right columns, are in increments of $0.03$, $0.06$, $0.05$ and $0.06$, respectively.}%The contours levels are $0.06, 0.12$ of the respective local rms value, except for (\textit{a,e}) for which they are $0.04,0.08$ times the local rms value.}%
  		\label{fig:spectra_2d_sp_cp_y15_g}
\end{figure}

Although $u'^*$ and $\overline{u'v'}^*$ within the canopy appear similar to those over smooth walls, there are some differences in the distribution of energy across different lengthscales in the flow, particularly in the region close to the wall. In order to examine this, we compare the spectral energy densities, at $y^* = 15$, for a smooth-wall and for case P1 in figure~\ref{fig:spectra_2d_sp_cp_y15_g}. This is the height roughly corresponding to the location at which the magnitude of the fluctuations peaks in smooth-wall flows \citep{Jimenez1999}. In global units, the energy is observed to be in larger wavelengths when compared to a smooth channel, especially in $\lambda_z$. In local scaling, however, there is a greater overlap of the regions with highest intensity, particularly for $E_{uu}$ and $E_{uv}$. In addition, the canopy case exhibits a concentration of energy at the canopy wavelengths and its harmonics. Note that the canopy spacing, for case P1, at $y^* \approx 15$ is reduced to $L_{x}^* = L_{z}^* \approx 100$,  while in global scaling it is $L_{x}^+ = L_{z}^+ \approx 200$. The increase in the energy in the canopy wavelengths is a reflection of the element-induced flow. The large streamwise scales, in turn, are damped by the presence of the canopy, which results in a reduction of their energy.
%\vspace{0.5cm}\includegraphics[scale=1.0,trim={0mm 2.5mm 0mm 0mm},clip]{images_posts/spectra_16x8_p_sm_y15.png}%

%\vspace{0.5cm} \includegraphics[scale=1.0,trim={0mm 0mm 0mm 0mm},clip]{images_posts/spectra_16x8_y_105.png}%
%\vspace{0.5cm}\hspace{-3mm} \includegraphics[scale=1,trim={0mm 0mm 0mm 0mm},clip]{images_posts/spectra_posts_y_250.png}%
\begin{figure}
		\vspace{0.5cm} \includegraphics[scale=1.0,trim={0mm 0mm 0mm 0mm},clip]{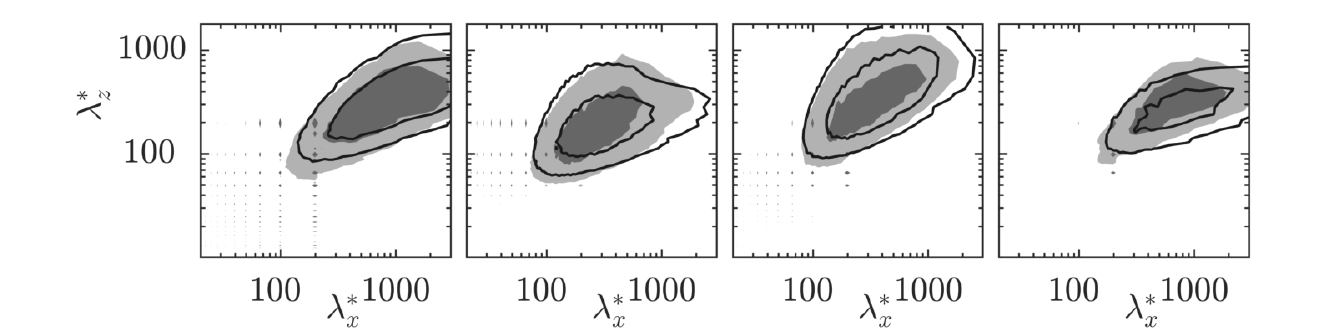}%
        \mylab{-11.4cm}{2.85cm}{(\textit{a})}%
  		\mylab{-8.7cm}{2.85cm}{(\textit{b})}%
  		\mylab{-6.0cm}{2.85cm}{(\textit{c})}%
  		\mylab{-3.3cm}{2.85cm}{(\textit{d})}%
  		\mylab{-10.8cm}{3.25cm}{$k_x k_z E_{uu}$}%
   		 \mylab{-8.1cm}{3.25cm}{$k_x k_z E_{vv}$}%
   		 \mylab{-5.40cm}{3.25cm}{$k_x k_z E_{ww}$}% 	
   		 \mylab{-2.70cm}{3.25cm}{$k_x k_z E_{uv}$}% 	
  		\caption{Pre-multiplied spectral energy densities for case P1 (filled contours), and case S (line contours) at $y^* = 105$, normalised by their respective $u^*$. The contours in ($a$--$d$) are in increments of 0.125, 0.06, 0.075 and 0.06, respectively.}% $(0.3, 0.03, 0.1, 0.06) \ u^{*2}$; $(0.15, 0.05, 0.06, 0.05)\ u^{*2}$; $(0.1, 0.04, 0.06, 0.04) \ u^{*2}$}%
   	     \label{fig:spectra_16x8_posts_y105}
   	     
		\vspace{0.5cm}\hspace{-3mm} \includegraphics[scale=1,trim={0mm 0mm 0mm 0mm},clip]{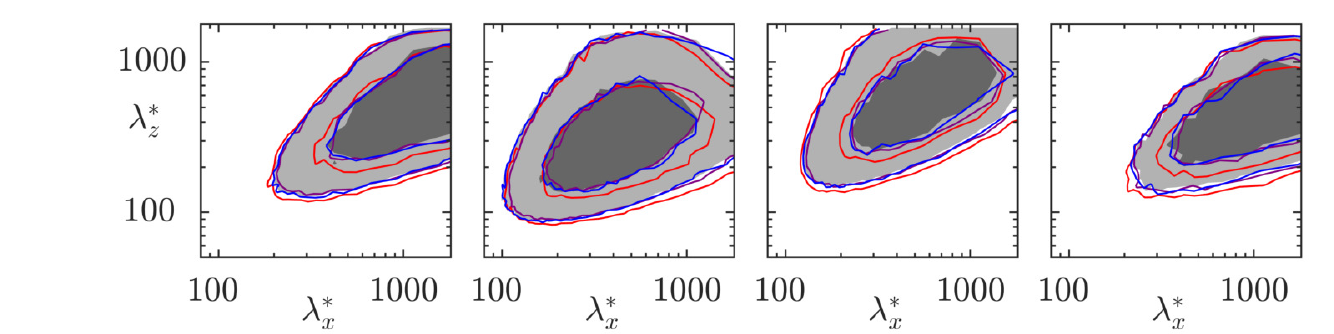}%
         \mylab{-11.4cm}{2.85cm}{(\textit{a})}%
  		\mylab{-8.5cm}{2.85cm}{(\textit{b})}%
  		\mylab{-5.65cm}{2.85cm}{(\textit{c})}%
  		\mylab{-2.75cm}{2.85cm}{(\textit{d})}%
  		\mylab{-10.8cm}{3.25cm}{$k_x k_z E_{uu}$}%
   		 \mylab{-7.9cm}{3.25cm}{$k_x k_z E_{vv}$}%
   		 \mylab{-5.10cm}{3.25cm}{$k_x k_z E_{ww}$}% 	
   		 \mylab{-2.20cm}{3.25cm}{$k_x k_z E_{uv}$}% 	
  		\caption{Pre-multiplied spectral energy densities at $y^+ = 250$. \protect\redline, case P0; \protect\violetline, case P1; and \protect\blueline, case P2, normalised by their respective $u_\tau$. Filled contours represent case S. The contours in ($a$--$d$) are in increments of $0.075$, $0.04$, $0.06$ and $0.03$, respectively.}% $(0.3, 0.03, 0.1, 0.06) \ u^{*2}$; $(0.15, 0.05, 0.06, 0.05)\ u^{*2}$; $(0.1, 0.04, 0.06, 0.04) \ u^{*2}$}%
   	     \label{fig:spectra_posts_y250}
\end{figure} 

\begin{figure}
	\centering
         \subfloat{%
%  		 \tikzsetnextfilename{Up_vs_y}
%  		\input{img_all/Up_vs_y.tex}
  		\includegraphics[scale=1]{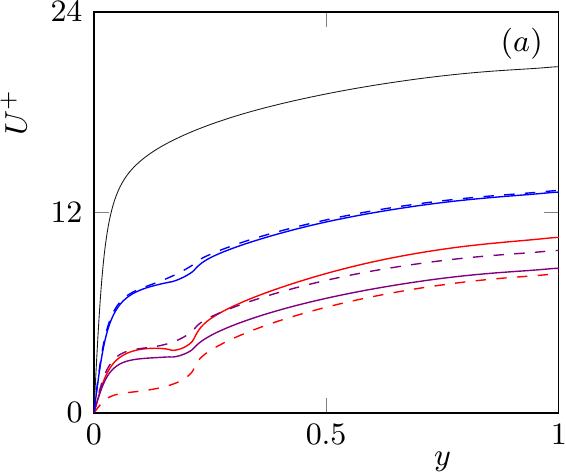}
  		}%  		
         \hspace{3mm} \subfloat{%
%  		 \tikzsetnextfilename{Up_vs_yp}
%  		\input{img_all/Up_vs_yp.tex}
  		\includegraphics[scale=1]{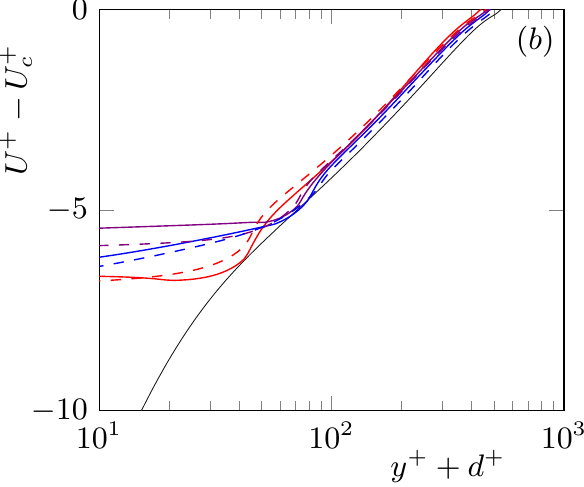}
  		}%
 \caption{{Mean velocity profiles, from the canopy-resolving simulations. Lines represent \protect\redline, case T1; \protect\violetline, case TP1; \protect\blueline, case T2; \protect\reddashed, case P0; \protect\violetdashed, case P1; \protect\bluedashed, case P2. The black lines represent the smooth-wall case, S. $U_c$ is the mean velocity at $y = \delta$.}}%
	    	\label{fig:mean_velocity_profiles}%
\end{figure} 

\begin{figure}
		 \includegraphics[scale=1.0,trim={0mm 6mm 0mm 0mm},clip]{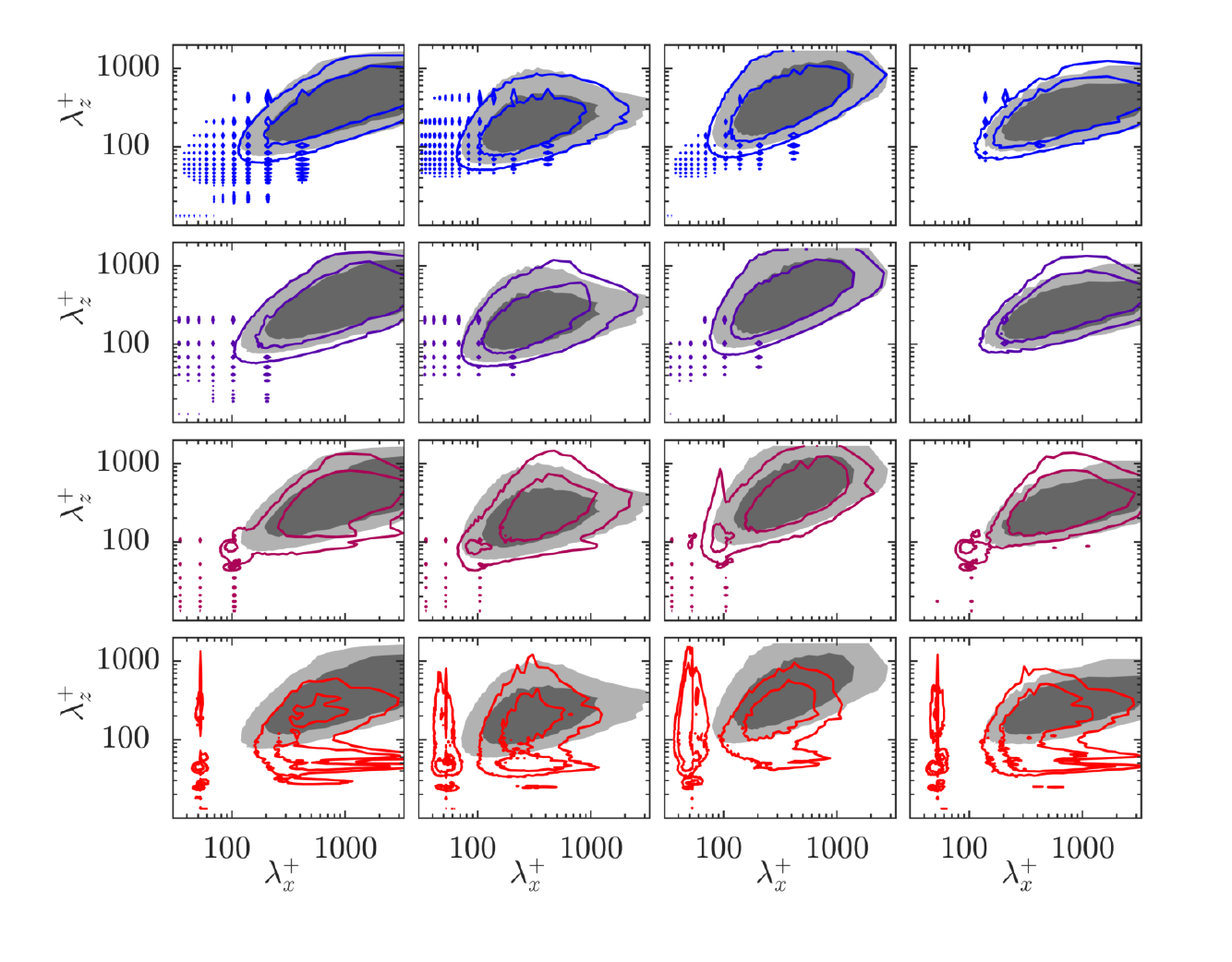}%
        \mylab{-11.45cm}{9.2cm}{(\textit{a})}%
  		\mylab{-8.7cm}{9.2cm}{(\textit{b})}%
  		\mylab{-6.0cm}{9.2cm}{(\textit{c})}%
  		\mylab{-3.3cm}{9.2cm}{(\textit{d})}%
         \mylab{-11.45cm}{7.0cm}{(\textit{e})}%
  		\mylab{-8.7cm}{7.0cm}{(\textit{f})}%
  		\mylab{-6.0cm}{7.0cm}{(\textit{g})}%
  		\mylab{-3.3cm}{7.0cm}{(\textit{h})}%
  		 \mylab{-11.45cm}{4.8cm}{(\textit{i})}%
  		\mylab{-8.7cm}{4.8cm}{(\textit{j})}%
  		\mylab{-6.0cm}{4.80cm}{(\textit{k})}%
  		\mylab{-3.3cm}{4.80cm}{(\textit{l})}%
  		 \mylab{-11.45cm}{2.65cm}{(\textit{m})}%
  		\mylab{-8.7cm}{2.65cm}{(\textit{n})}%
  		\mylab{-6.0cm}{2.65cm}{(\textit{o})}%
  		\mylab{-3.3cm}{2.65cm}{(\textit{p})}%
  		\mylab{-10.8cm}{9.6cm}{$k_x k_z E_{uu}$}%
   		 \mylab{-8.0cm}{9.6cm}{$k_x k_z E_{vv}$}%
   		 \mylab{-5.4cm}{9.6cm}{$k_x k_z E_{ww}$}% 	
   		 \mylab{-2.8cm}{9.6cm}{$k_x k_z E_{uv}$}% 	
  		\caption{{Pre-multiplied spectral energy densities at $y^* \approx 115$, normalised by their respective rms values. The line contours represent ($a$--$d$), case P2; ($e$--$h$), case P1; ($i$--$l$), case P0; ($m$--$p$), case PD. Filled contours represent case S. The contour increments in the leftmost to rightmost column are $0.029$, $0.048$, $0.045$ and $0.045$, respectively.}}% $(0.3, 0.03, 0.1, 0.06) \ u^{*2}$; $(0.15, 0.05, 0.06, 0.05)\ u^{*2}$; $(0.1, 0.04, 0.06, 0.04) \ u^{*2}$}%
   	     \label{fig:spectra_posts_32x16}
\end{figure} 

The differences in the energy distribution observed within the canopy eventually disappear above it. To illustrate this, figure~\ref{fig:spectra_16x8_posts_y105} portrays the spectra near the canopy tips, $y^* \approx 105$. Here, the concentration of energy in the canopy wavelengths and its harmonics is weak, and the smaller scales in the flow are smooth-wall like. There is, however, still a deficit of energy in large streamwise wavelengths compared to a smooth wall, associated with the damping of these scales by the canopy elements, as discussed in the previous paragraph. This effect diminishes away from the canopy, and the spectra are essentially smooth-wall-like for $y^+ \gtrsim 250$, as shown in figure~\ref{fig:spectra_posts_y250}, indicating that outer-layer similarity is recovered beyond this height. {Consequently, we can also conclude that this height marks the extent of the roughness sublayer of the canopies. The recovery of outer-layer similarity is also reflected in the mean-velocity profiles of the canopy flow simulations, portrayed in figure~\ref{fig:mean_velocity_profiles}, which exhibit logarithmic-law behaviour with a standard K\'arm\'an constant when shifted by a suitable displacement height, $d$ \citep{Jackson1981}.}

So far, we have mainly focussed on the results for case P1, with prismatic canopy elements with spacings $L_x^+ = L_z^+ \approx 200$. We now discuss the effect of the canopy element geometry and spacing. An increased sparsity results in an increase in the magnitude of both the full and background-turbulent velocity fluctuations, as shown in figure~\ref{fig:stats_resolved_gl_Posts}. In local scaling, however, the background turbulent fluctuations follow a similar trend to that observed for case P1. We observe that $u'^*$ and $\overline{u'v'}^*$ appear smooth-wall-like, while there is a relative increase in the magnitude of the cross fluctuations {compared to those over a smooth wall}. For the denser canopy of case P0, on the other hand, the fluctuations become less similar to those over smooth-walls. The streamwise fluctuations are damped more intensely within the canopy, and there are additional Reynolds shear stresses near the wall. Figure~\ref{fig:spectra_posts_32x16} shows that, compared to the sparser canopies, P0 has an accumulation of energy in streamwise wavelengths corresponding to the canopy harmonics but across a range of spanwise wavelengths. These regions of excess energy have also been noted by \citet{Abderrahaman2019}, who studied densely packed cuboidal roughness. They noted that these regions were an imprint of the large, background-turbulence scales modulating the smaller scale coherent flow generated by the roughness. This effect diminishes as the canopy element spacing is made larger than the energetic scales in the background-turbulence, as evidenced by the lack of these regions in the spectra of the sparser canopies portrayed in figures~\ref{fig:spectra_2d_sp_cp_y15_g}, \ref{fig:spectra_16x8_posts_y105} and \ref{fig:spectra_posts_32x16}. For case P2, the spectra are already close to smooth-wall-like near the canopy tips, suggesting that both the element-induced flow and the damping of large scales are already weak at this height.

\begin{figure}
	\centering
         \subfloat{%
%  		 \tikzsetnextfilename{urms_P_g_KH}
%  		\input{img_KH/u_c_g.tex}
  		\includegraphics[scale=1]{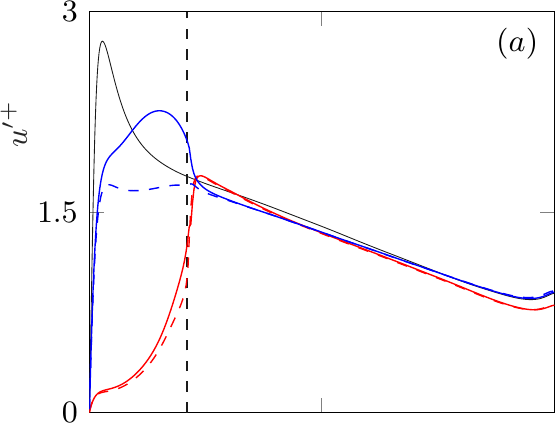}
  		}%  		
         \hspace{3mm}\subfloat{%
%  		 \tikzsetnextfilename{vrms_P_g_KH}
%  		\input{img_KH/v_c_g.tex}
  		\includegraphics[scale=1]{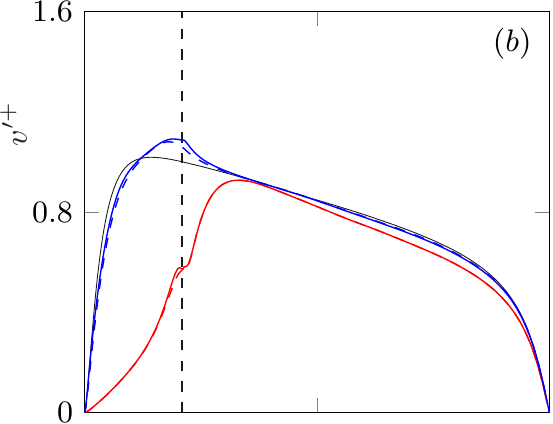}
  		}%
  		
  		\vspace{3mm}\hspace{1mm}\subfloat{%
% 		\tikzsetnextfilename{wrms_P_g_KH}
% 		\input{img_KH/w_c_g.tex}
 		\includegraphics[scale=1]{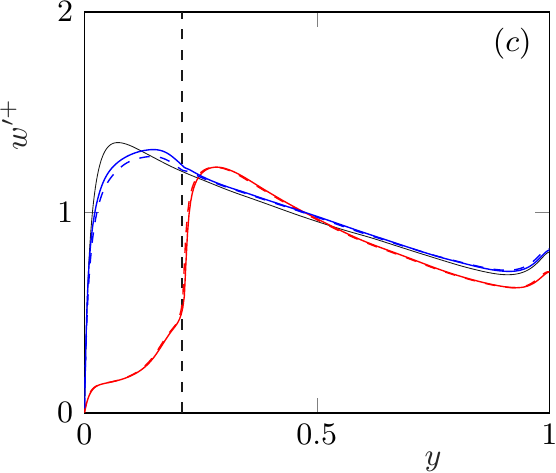}
 		}%		
 		\hspace{2.5mm}\subfloat{%
% 		\tikzsetnextfilename{uvrms_P_g_KH}
% 		\input{img_KH/Tau_c.tex}
 		\includegraphics[scale=1]{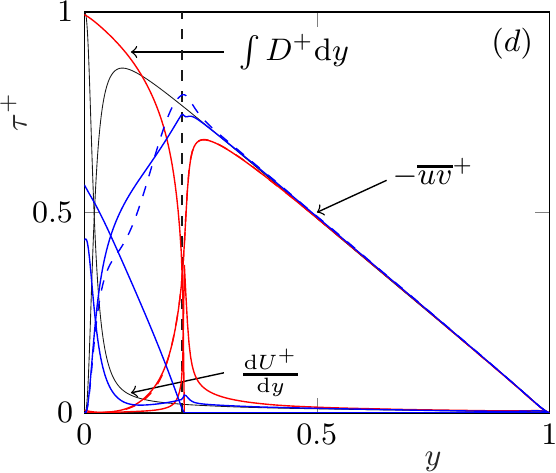}
 		}%
 		 \caption{{Rms velocity fluctuations and shear stresses scaled with the global friction velocity, $u_\tau$, from \protect\redline, case PD; and \protect\blueline, case P2. Solid lines represent the full velocity fluctuations and dashed lines represent the background-turbulence fluctuations. The vertical dashed line marks the location of the canopy tip plane. The black lines represent the smooth-wall case, S, for reference.}}%
	    	\label{fig:stats_resolved_g_KH}%
\end{figure}

\begin{figure}
	\centering
  		\vspace{3mm}\subfloat{%
  	    \hspace{-1mm}\includegraphics[scale=0.7,trim={8mm 5mm 15mm 2mm},clip]{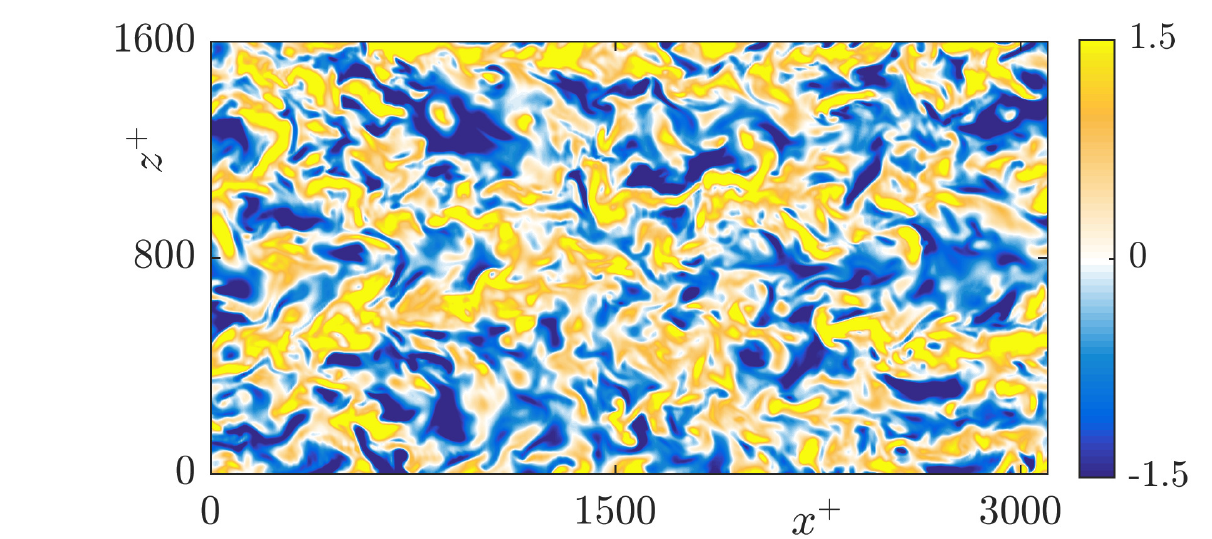}
  	     \mylab{-6.15cm}{3.3cm}{{(\textit{a})}}%
  		}%
 		
  		\subfloat{%
  	    \hspace{-1mm}\includegraphics[scale=0.7,trim={8mm 5mm 15mm 2mm},clip]{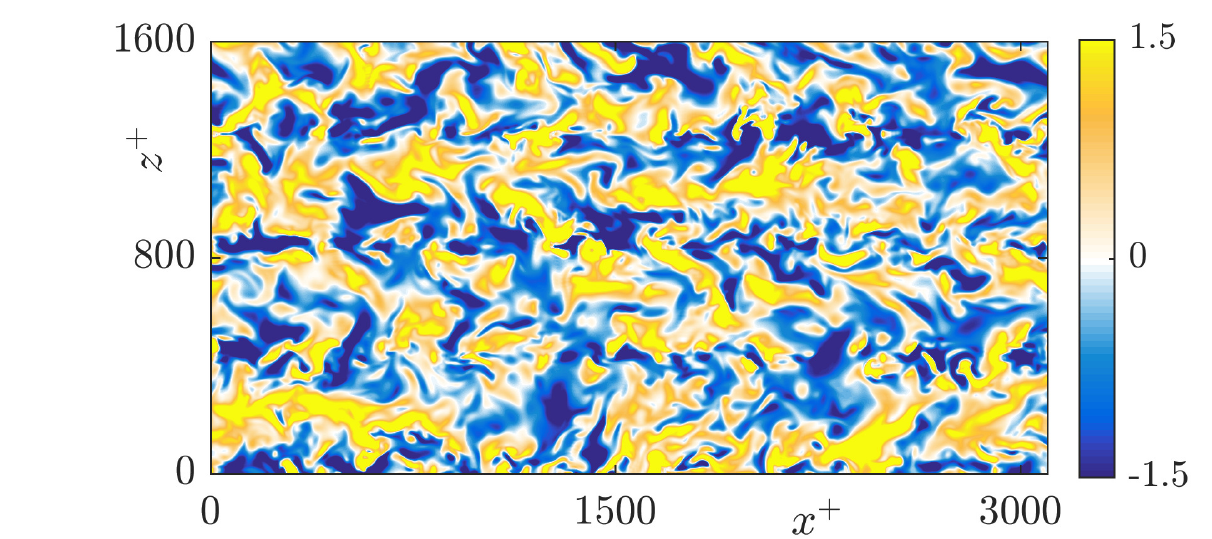}
  	     \mylab{-6.15cm}{3.3cm}{{(\textit{b})}}%
  		}%
  		
  		\subfloat{%
  	    \hspace{-1mm}\includegraphics[scale=0.7,trim={8mm 5mm 15mm 2mm},clip]{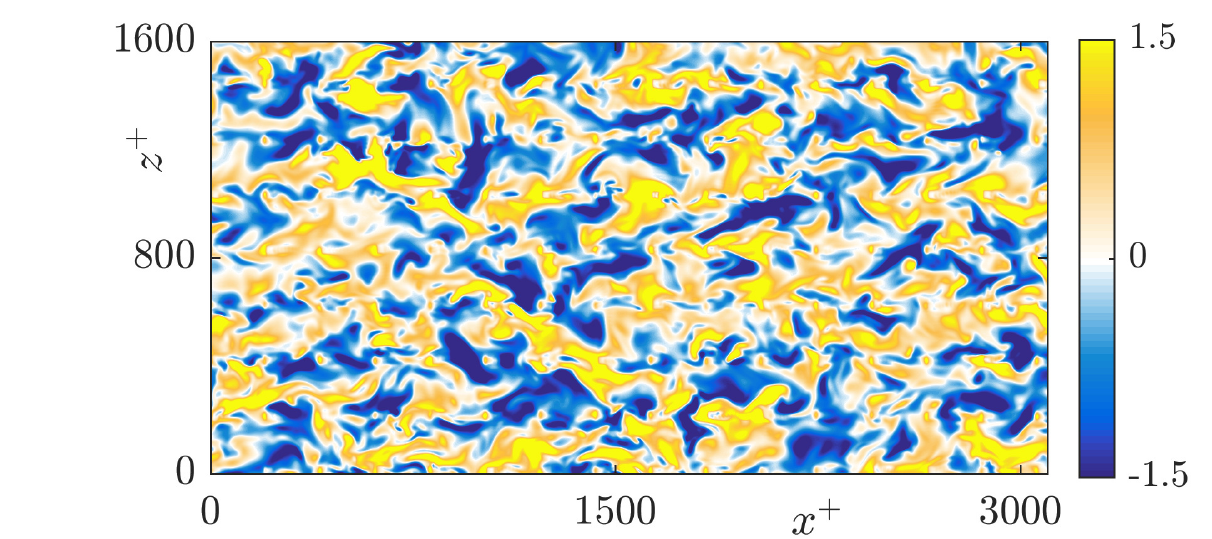}
  	      \mylab{-6.15cm}{3.3cm}{{(\textit{c})}}%
  		}%
 		
  		\subfloat{%
  	   \hspace{-2mm} \includegraphics[scale=0.7,trim={8mm 5mm 15mm 2mm},clip]{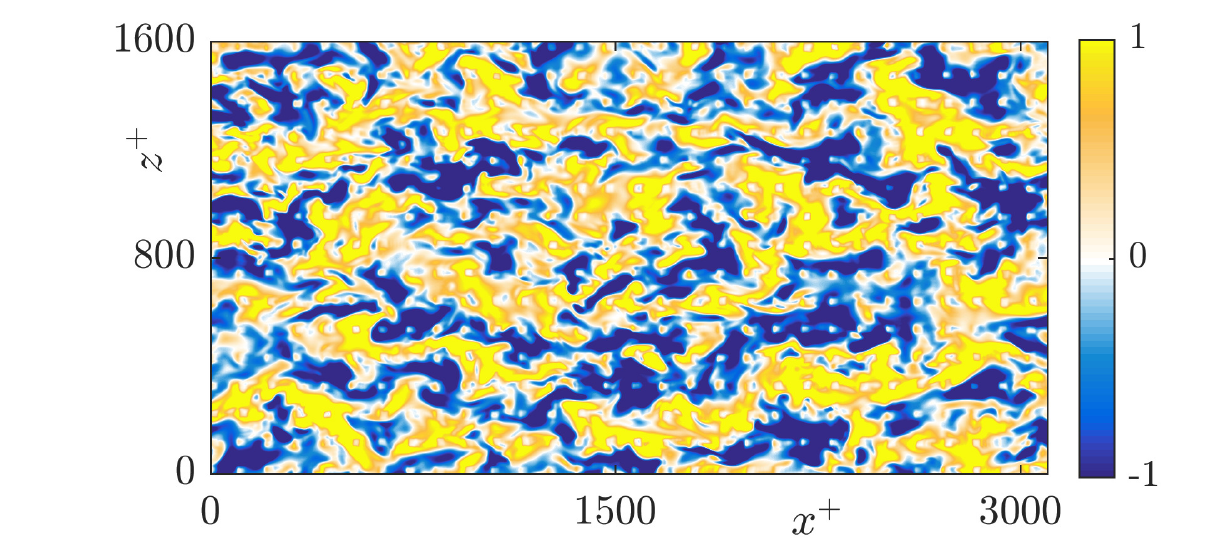}
  	     \mylab{-6.15cm}{3.3cm}{{(\textit{d})}}%
  		}%
 		
  		\subfloat{%
  	    \hspace{-1mm}\includegraphics[scale=0.7,trim={8mm 0mm 15mm 2mm},clip]{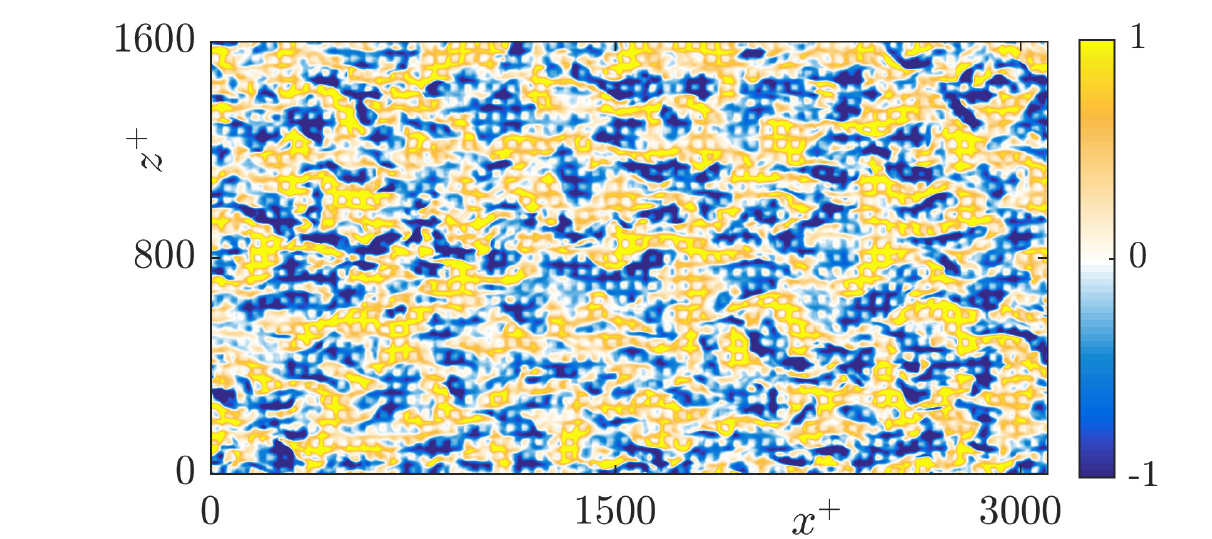}
  	      \mylab{-6.15cm}{3.65cm}{{(\textit{e})}}%
  		}%
 		\caption{{Instantaneous realisations of the wall normal velocity at $y^+ \approx 120$, normalised by $u_\tau$. The panels from ($a$) to ($e$) represent cases S, P2, P1, P0 and PD. The clearest and darkest colours in ($a$) to ($e$) represent intensities of $\pm(1.5, 1.5, 1.5, 1, 1)$, respectively.}}%
   	     \label{fig:v_snap_G}
\end{figure}

{The results discussed above suggest that there is a progressive departure from smooth-wall-like behaviour in canopy flows as the element spacing is reduced, which is consistent with the observations of \citet{Poggi2004} and \citet{Huang2009}. If the element spacing was reduced even further, eventually we would expect the complete breakdown of smooth-like dynamics within the canopy. In the resulting dense canopy, the flow near the canopy tips would be characterised by the Kelvin--Helmholtz-like, mixing-layer instability \citep{Raupach1996,Finnigan2000,Poggi2004,Nepf2012}. To investigate this effect, we have conducted an additional simulation of the prismatic post canopies in an even denser arrangement, case PD. In order to contrast the flow characteristics of sparse canopies with those of dense ones, we now compare the results from the simulation PD to that of the sparsest canopy studied, case P2. Deep within the dense canopy of case PD, the viscous and the Reynolds shear stresses are negligible and the drag stress dominates, as shown in figure~\ref{fig:stats_resolved_g_KH}($d$). While for the sparse canopy of case P2, the magnitude of the velocity fluctuations near the canopy tips is similar to that over a smooth-wall, for the dense canopy of case PD their magnitude is considerably reduced. Furthermore, the magnitude of the element-induced streamwise fluctuations in the dense canopy is negligible compared to that of the sparse canopy, where it is observed to constitute up to $30\%$ of the total fluctuations, similar to the observations of \citet{Poggi2008}. Previous studies have noted the formation of Kelvin--Helmholtz-like instabilities near the canopy tips in dense canopies \citep{Finnigan2000,Poggi2004,Nepf2012}. When present, these instabilities leave a distinct footprint in $E_{vv}$ and $E_{uv}$, causing an increase in energy in a narrow range of streamwise wavelengths and for large spanwise wavelengths \citep{Garcia-Mayoral2011,GG2018,Abderrahaman2019}. For the canopies of cases P0, P1 and P2, such footprint is not observed in the spectral energy densities, portrayed in figure~\ref{fig:spectra_posts_32x16}. In the energy densities of the wall-normal velocity for the dense canopy, case PD, we observe some concentration of energy in a range of streamwise wavelengths, $\lambda_x^+ \in 200$--$500$, consistent with the presence of a Kelvin--Helmholtz-like instability. The gradual breakdown of the smooth-wall-like behaviour of flows over canopies with decreasing element spacing can also be observed in instantaneous realisations of the wall-normal velocity, portrayed in figure~\ref{fig:v_snap_G}. Dense canopies are not the focus of the present study so the results from case PD are not discussed extensively here.} {Increasing the canopy density yet further can result in a more distinct imprint of the Kelvin--Helmholtz-like instability near the canopy tips, as discussed in \citet{Poggi2004} and \citet{Sharma2019dense}. Further details about the formation and development of this instability over dense canopies can be found in \citet{Raupach1996,Ghisalberti2002,Finnigan2009} and \citet{Bailey2016}.}

%\begin{figure}
%	\centering
%  		\vspace{3mm}\subfloat{%
%  	    \hspace{-1mm}\includegraphics[scale=0.7,trim={8mm 5mm 15mm 2mm},clip]{sparse_canopy_snapshots/vsnap_y100_smooth.png}
%  	     \mylab{-6.15cm}{3.3cm}{{(\textit{a})}}%
%  		}%
% 		
%  		\subfloat{%
%  	    \hspace{-1mm}\includegraphics[scale=0.7,trim={8mm 5mm 15mm 2mm},clip]{sparse_canopy_snapshots/vsnap_y100_8x4.png}
%  	     \mylab{-6.15cm}{3.3cm}{{(\textit{b})}}%
%  		}%
%  		
%  		\subfloat{%
%  	    \hspace{-1mm}\includegraphics[scale=0.7,trim={8mm 5mm 15mm 2mm},clip]{sparse_canopy_snapshots/vsnap_y100_16x8.png}
%  	      \mylab{-6.15cm}{3.3cm}{{(\textit{c})}}%
%  		}%
% 		
%  		\subfloat{%
%  	   \hspace{-2mm} \includegraphics[scale=0.7,trim={8mm 5mm 15mm 2mm},clip]{sparse_canopy_snapshots/vsnap_y100_32x16.png}
%  	     \mylab{-6.15cm}{3.3cm}{{(\textit{d})}}%
%  		}%
% 		
%  		\subfloat{%
%  	    \hspace{-1mm}\includegraphics[scale=0.7,trim={8mm 0mm 15mm 2mm},clip]{sparse_canopy_snapshots/vsnap_y100_64x32.png}
%  	      \mylab{-6.15cm}{3.65cm}{{(\textit{e})}}%
%  		}%
% 		\caption{{Instantaneous realisations of the wall normal velocity at $y^+ \approx 120$, normalised by $u_\tau$. The panels from ($a$) to ($e$) represent cases S, P2, P1, P0 and PD. The clearest and darkest colours in ($a$) to ($e$) represent intensities of $\pm(1.5, 1.5, 1.5, 1, 1)$, respectively.}}%
%   	     \label{fig:v_snap_G}
%\end{figure}	

\begin{figure}
	\centering
  		\subfloat{%
%  		 \tikzsetnextfilename{urms_PH_g}
%  		\input{images_posts/u_PH_g.tex}
  		\includegraphics[scale=1]{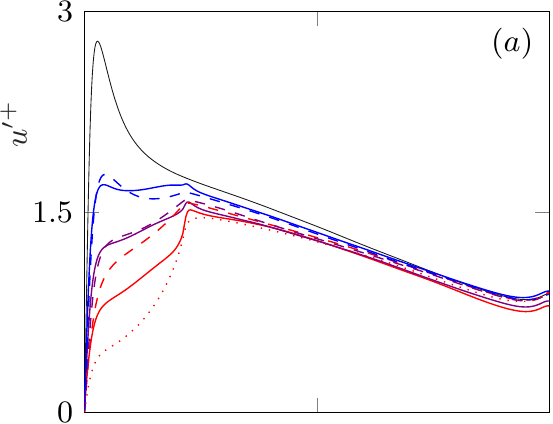}
  		}%
         \hspace{3mm}\subfloat{%
%  		 \tikzsetnextfilename{vrms_PH_g}
%  		\input{images_posts/v_PH_g.tex}
  		\includegraphics[scale=1]{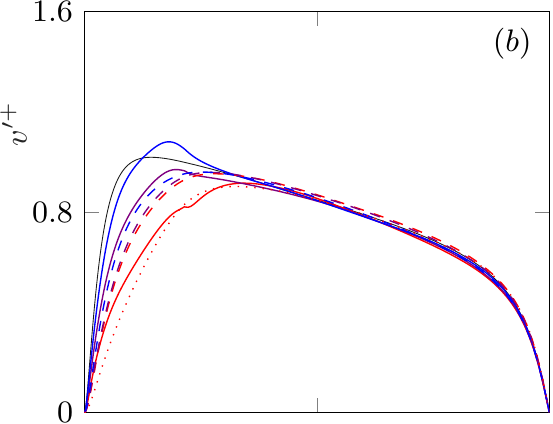}
  		}%
  	
        \vspace{3mm}\hspace{0.4mm}\subfloat{%
% 		\tikzsetnextfilename{wrms_PH_g}
% 		\input{images_posts/w_PH_g.tex}
 		\includegraphics[scale=1]{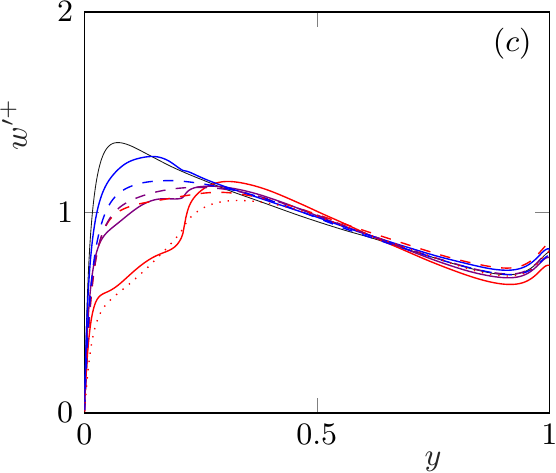}
 		}%	 		 		
 		\hspace{2.7mm}\subfloat{%
% 		\tikzsetnextfilename{uvrms_PH_g}
% 		\input{images_posts/Re_stress_PH_g.tex}
 		\includegraphics[scale=1]{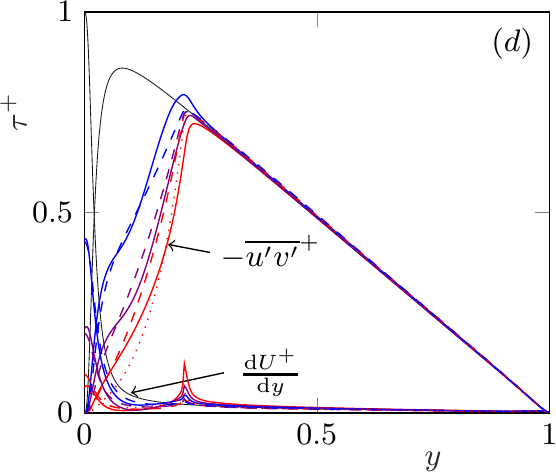}
 		}%
  	 \caption{Background-turbulence rms velocity fluctuations and shear stresses scaled with the global friction velocity, $u_\tau$, of the canopy-resolving and mean-only/homogeneous drag simulations. The lines represent \protect\redline, case P0; \protect\reddashed, case P0-H0; \protect\reddotted, case P0-H; \protect\violetline, case P1; \protect\violetdashed, case P1-H0; \protect\blueline, case P2; and \protect\bluedashed, case P2-H0. The black lines represent the smooth-wall case, S.}%
	    	\label{fig:stats_models_g_Posts}	
\end{figure} 
\begin{figure}
	\centering
  		\subfloat{%
%  		 \tikzsetnextfilename{urms_TH_g}
%  		 \input{images_Ts/u_TH_g.tex}
  		 \includegraphics[scale=1]{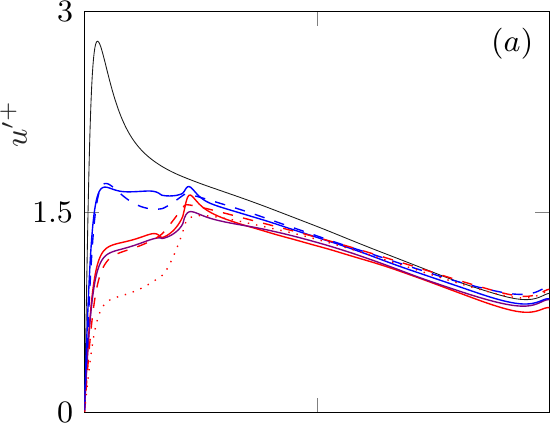}
  		}%
  	    \hspace{3mm}\subfloat{%
%  		 \tikzsetnextfilename{vrms_TH_g}
%  		 \input{images_Ts/v_TH_g.tex}
  		 \includegraphics[scale=1]{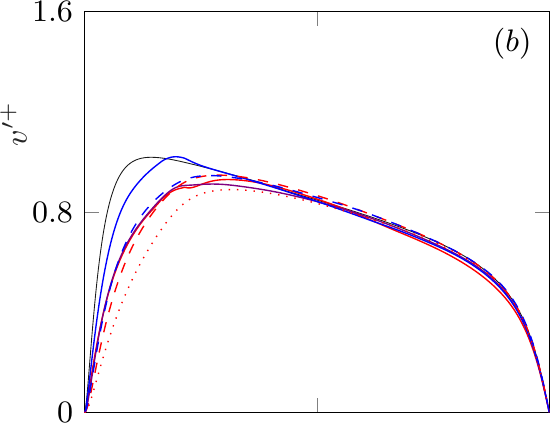}
  		}%
  	
        \vspace{3mm}\hspace{0.4mm}\subfloat{%
% 		\tikzsetnextfilename{wrms_TH_g}
% 		\input{images_Ts/w_TH_g.tex}
 		\includegraphics[scale=1]{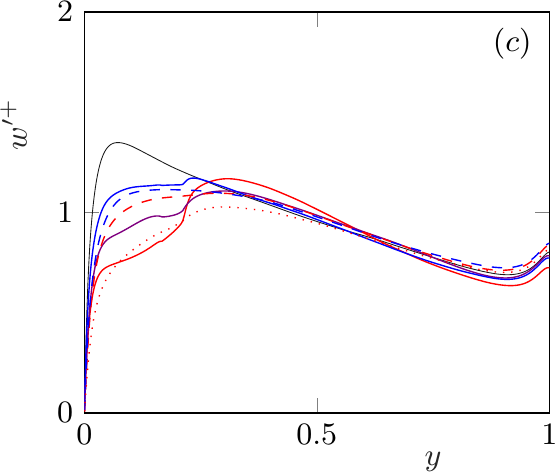}
 		}%	
 	    \hspace{2.7mm}\subfloat{%
% 		\tikzsetnextfilename{uvrms_TH_g}
% 		\input{images_Ts/Re_stress_TH_g.tex}
 		\includegraphics[scale=1]{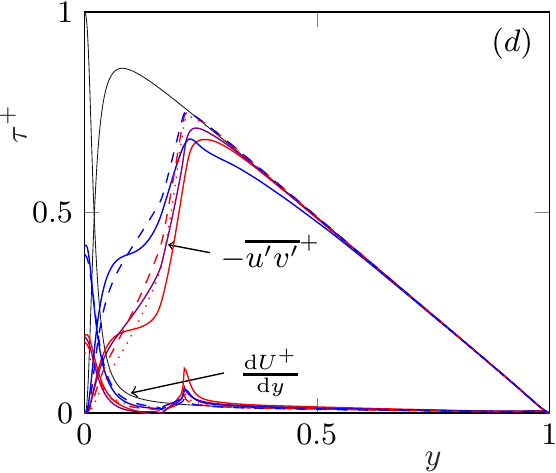}
 		}%
  	 \caption{Background-turbulence rms velocity fluctuations and shear stresses scaled with the global friction velocity, $u_\tau$, of the canopy-resolving and mean-only/homogeneous drag simulations. The lines represent \protect\redline, case T1; \protect\reddashed, case T1-H0; \protect\reddotted, case T1-H; \protect\violetline, case TP1; \protect\blueline, case T2; and \protect\bluedashed, case T2-H0. The black lines represent the smooth-wall case, S.}%
	    	\label{fig:stats_models_g_Ts}	
\end{figure}

\section{Simulations with artificial forcing} \label{sec:results_models}
The results discussed so far suggest that sparse canopies affect their surrounding flow through two mechanisms, an element-induced flow, and a change in the local scale for the background-turbulence fluctuations. With respect to the second mechanism, the effect of the canopy elements would be indirect, through modifying the mean-velocity profile and thus the local stress, $\tau_f$. The latter would, in turn, set the scale for the fluctuations. If this is the case, applying the mean drag produced by the canopy on the mean flow alone should capture the essential effects of the canopy on the background-turbulence. We test this in the simulations labelled with the suffix `-H0'. For the cases P0 and T1/TP1, we also compare the mean-only-drag simulations with conventional, homogeneous drag simulations labelled with the suffix `-H'. {These could be expected to be better representations when the canopy is dense enough for all the turbulent scales to perceive it in a homogenised fashion}. Note that simulations T1-H0 and T1-H correspond to both the permeable and impermeable canopies of T1 and TP1, as they have similar net drags and drag coefficients.  %To this end, in this section we compare a conventional, homogeneous drag model, with a mean-only drag model. %In the latter case, we also discuss the results obtained from fixing the mean velocity profile and obtaining the drag force a posteriori. In addition, we also explore a model where the mean-only drag is distributed in a low order representation of the canopy layout.

The streamwise fluctuations and the Reynolds shear stresses of the mean-only-drag simulations are in good agreement with the corresponding background-turbulence fluctuations from the resolved canopies, except for case P0, as shown in figures~\ref{fig:stats_models_g_Posts} and \ref{fig:stats_models_g_Ts}. For the sparsest canopies, cases P2 and T2, the cross-flow fluctuations, particularly in the wall-normal direction, are slightly larger than their mean-only-drag counterparts. A likely reason for this discrepancy is the presence of an unsteady element-induced flow for these canopies, whose contribution cannot be filtered out by the conventional triple-decomposition technique that we have used. The fluctuating velocities scaled by the local friction velocities for the drag-force simulations are provided in appendix~\ref{appA} for reference.

For case P0, a homogeneous drag provides a better representation of the cross-flow fluctuations than the mean-only drag, and the streamwise fluctuations are not well represented by either forcing method. For this case, there is significant interaction between the element-induced flow and the background turbulence, as discussed in \S\ref{sec:resolved_canopy}. Thus, it is not surprising that neither the mean-only drag nor the homogeneous drag is able to capture the full effect of this canopy on the background turbulence.

 \begin{figure}
		\centering
         \includegraphics[scale=1.0,trim={0mm 2mm 0mm 0mm},clip]{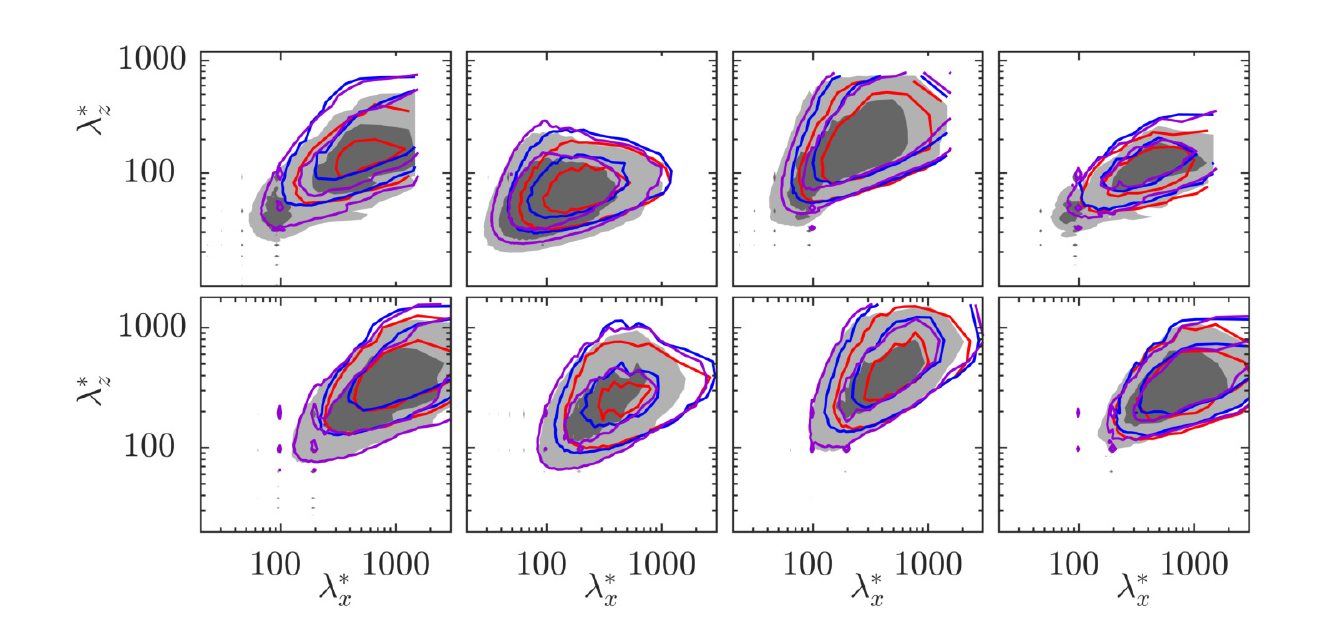}%
        \mylab{-11.45cm}{5.3cm}{(\textit{a})}%
  		\mylab{-8.7cm}{5.3cm}{(\textit{b})}%
  		\mylab{-6.0cm}{5.3cm}{(\textit{c})}%
  		\mylab{-3.3cm}{5.3cm}{(\textit{d})}%
         \mylab{-11.45cm}{2.80cm}{(\textit{e})}%
  		\mylab{-8.7cm}{2.80cm}{(\textit{f})}%
  		\mylab{-6.0cm}{2.80cm}{(\textit{g})}%
  		\mylab{-3.3cm}{2.80cm}{(\textit{h})}%
  		\mylab{-10.8cm}{5.7cm}{$k_x k_z E_{uu}$}%
   		 \mylab{-8.0cm}{5.7cm}{$k_x k_z E_{vv}$}%
   		 \mylab{-5.4cm}{5.7cm}{$k_x k_z E_{ww}$}% 	
   		 \mylab{-2.8cm}{5.7cm}{$k_x k_z E_{uv}$}% 	
			\caption{Pre-multiplied spectral energy densities at (\textit{a}--\textit{d}), $y^* = 15$ and (\textit{e}--\textit{h}), $y^* = 105$, normalised by their respective $u^*$. Filled contours represent case TP1; \protect\blueline, case T1-H0; \protect\newredline, case T1-H; and \protect\violetline, case TP1-L. The contours in ($a$--$h$) are in increments of $0.3$, $0.075$, $0.175$, $0.075$, $0.12$, $0.06$, $0.075$ and $0.05$, respectively.}%
 	\label{fig:spectra_2D_sp_md}		
 \end{figure}

For the sparser canopies of T1 and TP1, compared to a mean-only drag, the homogeneous drag tends to overdamp the fluctuations within the canopy, particularly in the streamwise direction, as can be observed in figure~\ref{fig:stats_models_g_Ts}($a$). The excessive damping of fluctuations by a homogeneous drag, in comparison to a resolved canopy, was also noted by \citet{Yue2007} and \citet{Bailey2013}. Figure~\ref{fig:spectra_2D_sp_md} shows that this decrease in the intensity of fluctuations within the canopy is mainly a result of damping of the smaller streamwise wavelengths in the flow, $\lambda_x^* \lesssim 200$. The homogeneous drag simulation, T1-H, reproduces well the larger scales of the resolved canopy simulation, TP1. This suggests that scales much larger than the canopy spacing still perceive the canopy as homogeneous. Using the mean-only drag recovers some of the smaller streamwise scales, but it does not act directly on the larger scales as the actual canopy does. As the element spacing is increased, the range of scales affected in a homogenised fashion is shifted to larger scales, so that the energetic turbulent scales become less damped. This is consistent with the results portrayed in figures~\ref{fig:spectra_posts_32x16}, which show that, near the canopy tips, the dense canopy P0 damps the energy at $\lambda_x^* \sim 1000-2000$, compared to smooth walls, while the sparse canopy P2 leaves these scales relatively undisturbed.

 \begin{figure}
		\centering
           % \vspace{}
%  		 \tikzsetnextfilename{reduced_order_cd}
% 			\input{Images_new/reduced_order_drag_coeff.tex}
 			\includegraphics[scale=1]{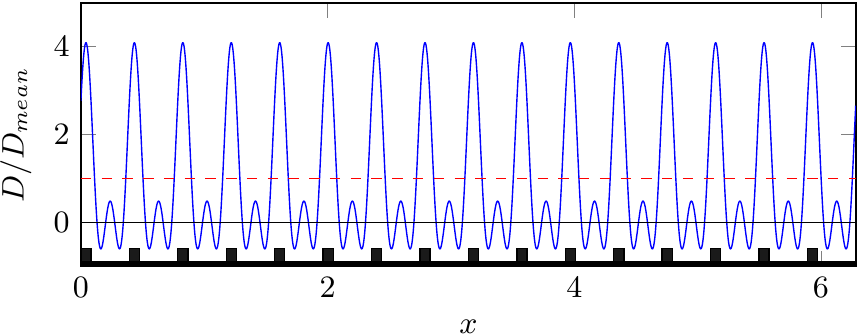}
    		\caption{Drag force distribution in the streamwise direction in a plane passing through the canopy heads for case TP1-L (blue); \protect\reddashed, distribution of the mean-only drag force, as in case T1-H0. The location of the canopy elements is sketched in grey at the bottom of the figure.}
	    	\label{fig:canopy_phys_fou}	
 \end{figure}  

\begin{figure}
	\centering
		\subfloat{%
%  		 \tikzsetnextfilename{urms_utg_yg_F}
%  		\input{images_Ts_temp/u_THF_g.tex}
  		\includegraphics[scale=1,trim={0mm 3mm 0mm 0mm},clip]{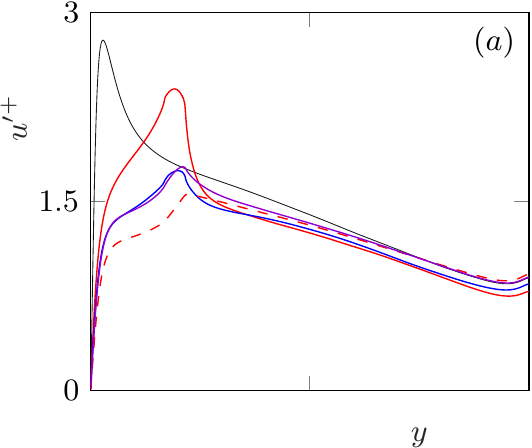}
  		}%
  		\hspace{3mm}\subfloat{%
%  		 \tikzsetnextfilename{vrms_utg_yg_F}
%  			\input{images_Ts_temp/v_THF_g.tex}
  			\includegraphics[scale=1,trim={0mm 3mm 0mm 0mm},clip]{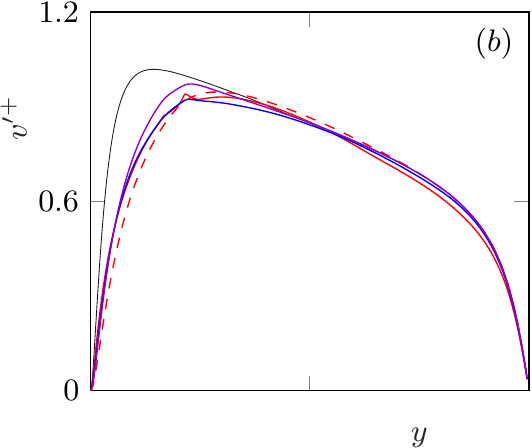}
  		}%
  		
  		\vspace{2mm}\hspace{0.5mm}\subfloat{%
% 		\tikzsetnextfilename{wrms_utg_yg_F}
% 			\input{images_Ts_temp/w_THF_g.tex}
 			\includegraphics[scale=1]{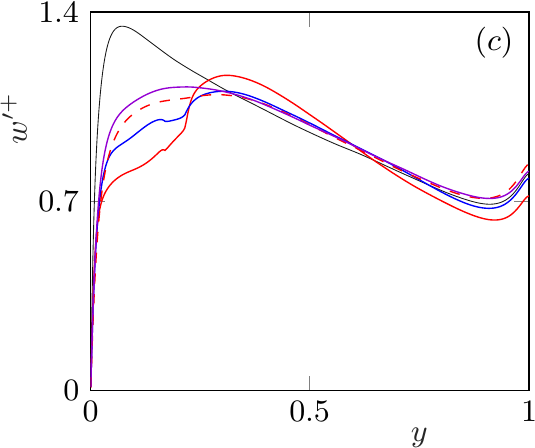}
 		}%
 		\hspace{3.5mm}\subfloat{%
% 		\tikzsetnextfilename{uvrms_utg_yg_F}
% 		\input{images_Ts_temp/Tau_THF.tex}
 		\includegraphics[scale=1]{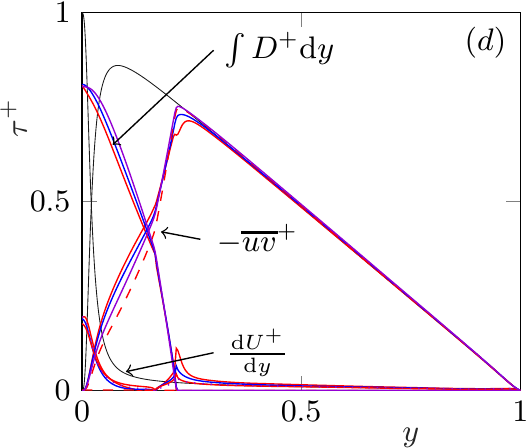}
 		}%
 		 \caption{Rms fluctuations and shear stresses of the full flow, scaled with the global friction velocity $u_\tau$. \protect\blueline, case TP1; \protect\redline, case T1; \protect\reddashed, case TP1-H0; \protect\violetline case TP1-L. The black lines represent case S.}%
 		  	\label{fig:stats_TDF_TD}%
\end{figure}

The accumulation of energy in the lengthscales of the order of the canopy wavelengths and its harmonics observed in the canopy-resolving simulations requires a discrete representation of the canopy elements. Hence, it cannot be captured by either the mean-only- or the homogeneous-drag approaches. To introduce information of the canopy layout in the model, we distribute the drag calculated from the mean flow into a reduced-order representation of the canopy elements in case TP1-L. The representation consists of a truncation in Fourier space in $x$ and $z$, of the actual layout. The procedure is illustrated in figure~\ref{fig:canopy_phys_fou} by the streamwise distribution of the drag force used by this model. In addition to capturing the local scaling of the flow, discussed in \S\ref{sec:results_models}, this model is also able to represent the concentration of energy in the canopy scales for case TP1, as observed in figure~\ref{fig:spectra_2D_sp_md}. This is reflected by the collapse of the rms fluctuations of the full velocity components of TP1-L and TP1, as shown in figure~\ref{fig:stats_TDF_TD}. The magnitude of spanwise velocity fluctuations within the canopy of TP1-L is slightly larger than TP1, likely due to the fact that TP1-L does not apply any form of spanwise drag force. The drag force, although only applied in the streamwise direction, is also able to reproduce the canopy harmonics in the spectra of $E_{vv}$ and $E_{ww}$, which are caused by the deflection of the streamwise flow around the canopy elements as a result of continuity. The large scales in the flow are similar to those in the mean-only drag, as the drag in this case also does not act on these scales directly. This method, however, is only able to capture the weak coherent flow generated by the permeable canopy, and still under-predicts the full streamwise velocity fluctuations of the impermeable canopy.
 \section{Conclusions} \label{sec:conclusions}
In the present work, we have studied turbulent flows within and over sparse canopies. Two different canopy element geometries have been studied, each for various different element spacings. We have also compared canopies with permeable and impermeable elements in the same arrangement. {The effect of the Reynolds number has also been examined by comparing the results from simulations of the same canopy in both inner and outer scaling at $\Rey_\tau \approx 520$ and $1000$}. The flow was decomposed into an element-induced component and a background-turbulence component. It was found that although the element-induced flow in the permeable and impermeable canopies studied here differ, the background turbulence was essentially the same. A new scaling for the background-turbulence fluctuations within sparse canopies was proposed. This scaling uses the friction velocity based on the local sum at each height of the viscous and Reynolds shear stresses, $\tau_f$, rather than the conventional friction velocity, based on the net drag. When scaled with the proposed local friction velocity, the background-turbulence fluctuations and the viscous and Reynolds shear stresses appear more smooth-wall-like, compared to when conventional total-drag scaling is used. This suggests that the sparse canopy acts in a large part on the background turbulence through a change in the local scale, rather than through a direct interaction with the canopy elements. Based on the proposed scaling, we investigated the extent to which a drag force acting only on the mean flow captures the effect of the canopy on the background turbulence. The mean-only drag directly modifies the mean flow alone, which in turn sets $\tau_f$ and, hence, the scale for the fluctuations. We show that the mean-only drag is able to capture the background-turbulence fluctuations within the canopies better than a conventional, homogeneous drag. Neither approach is, however, sufficient to capture the element-induced flow. The latter can be partially recovered by redistributing the mean-only drag in a low-order representation of the canopy. 

\bigskip
A.S. was supported by an award from the Cambridge Commonwealth, European and International Trust. Computational resources were provided by the	 ``Cambridge Service for Data Driven Discovery" operated by the University of Cambridge Research Computing Service and funded by EPSRC Tier-2 grant EP/P020259/1. Additional computing time was also provided by PRACE DECI-15. This work was also partly supported by the European Research Council through the Third Multiflow Summer Workshop. We would also like to thank Prof. H. Nepf for her helpful feedback on an earlier version of this manuscript. The authors report no conflicts of interest.

\appendix
\section{Turbulence statistics in local scaling}\label{appA}
\begin{figure}
	\centering
  		\subfloat{%
%  		 \tikzsetnextfilename{urms_Re_g2}
%  		\input{img_Re2/u_c_g_g2.tex}
  		\includegraphics[scale=1]{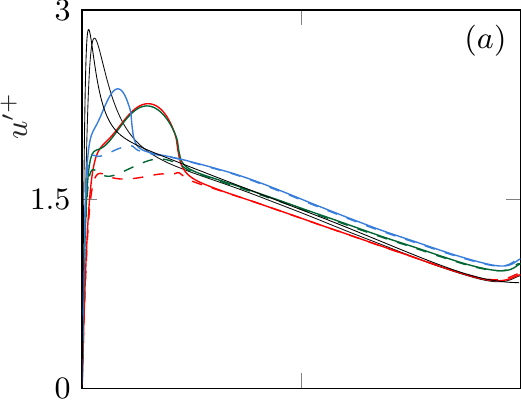}
  		}%
  		\hspace{3mm}\subfloat{%
%  		 \tikzsetnextfilename{urms_Re_l2}
%  		\input{img_Re2/u_c_l_l.tex}
  		\includegraphics[scale=1]{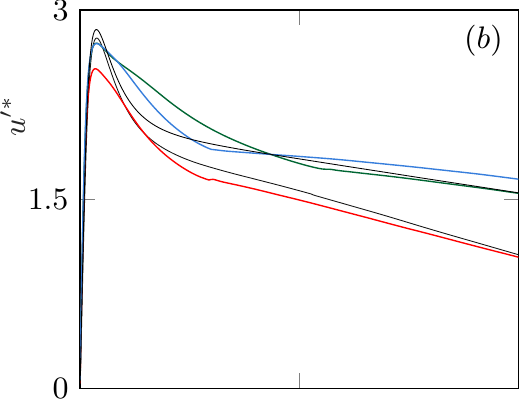}
  		}%
  		
  	    \vspace{0mm}\subfloat{%
%  		 \tikzsetnextfilename{vrms_Re_g2}
%  	       \input{img_Re2/v_c_g_g2.tex}
  	       \includegraphics[scale=1]{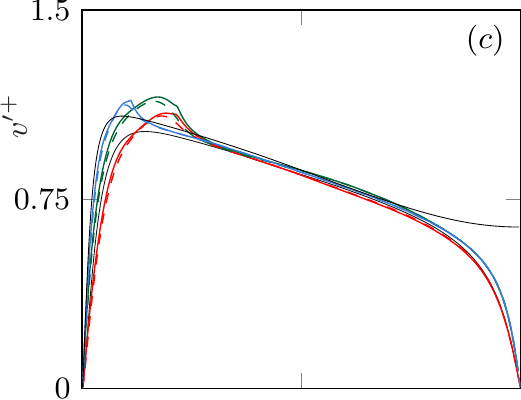}
  		}%
  		\hspace{3mm}\subfloat{%
%  		 \tikzsetnextfilename{vrms_Re_l2}
%  		\input{img_Re2/v_c_l_l.tex}
  		\includegraphics[scale=1]{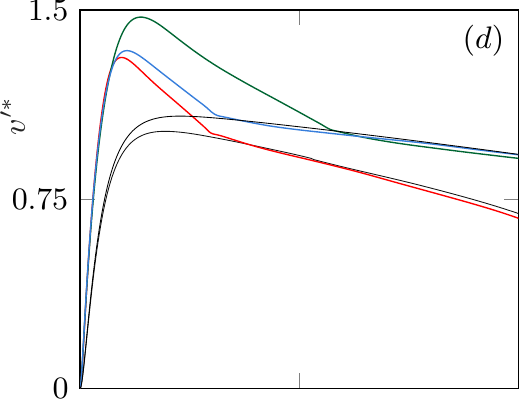}
  		}%
  		
  	  \vspace{0mm}\subfloat{%
% 		\tikzsetnextfilename{wrms_Re_g2}
% 			\input{img_Re2/w_c_g_g2.tex}
 			\includegraphics[scale=1]{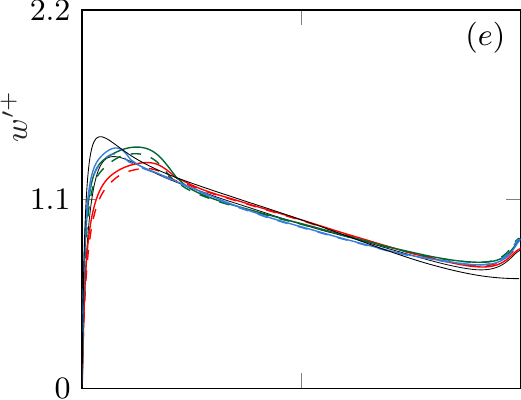}
 		}%
        \hspace{3mm}\subfloat{%
% 		\tikzsetnextfilename{wrms_Re_l2}
% 		\input{img_Re2/w_c_l_l.tex}
 		\includegraphics[scale=1]{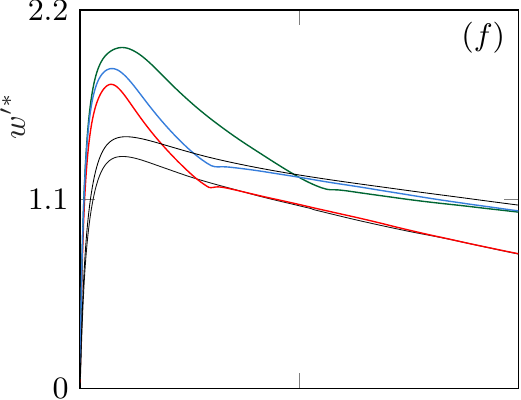}
 		}%		
 		
 		\vspace{3.5mm}\hspace{1.5mm}\subfloat{%
% 		\hspace{-1mm}\tikzsetnextfilename{uvrms_Re_g2}
% 			\input{img_Re2/Tau_c_g_g2.tex}
 			\includegraphics[scale=1]{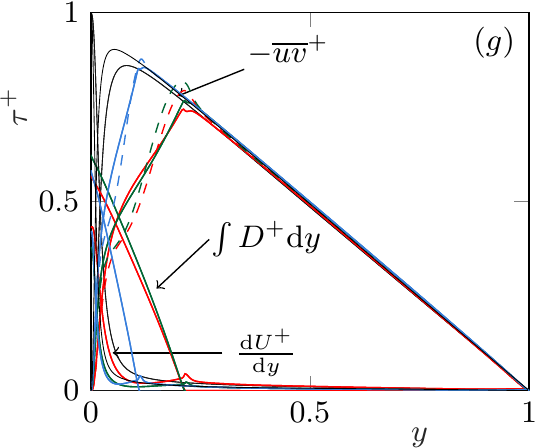}
 		}%
        \hspace{2.7mm}\subfloat{%
% 	   \hspace{-0.5mm}\tikzsetnextfilename{uvrms_Re_l2}
% 		\input{img_Re2/Re_stress_c_l_l.tex}
 		\includegraphics[scale=1]{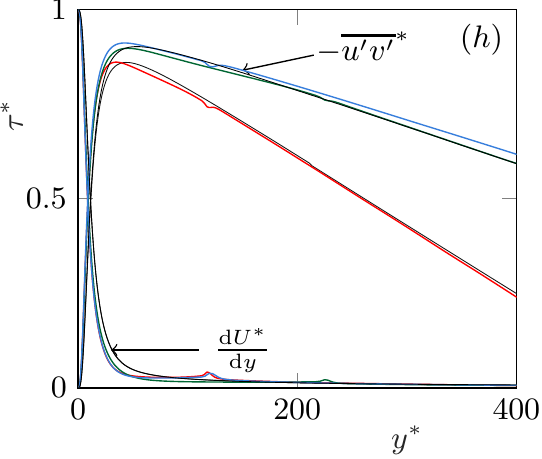}
 		}%
  	 \caption{{Rms velocity fluctuations and shear stresses scaled with the global friction velocity, $u_\tau$,  in the left column and with the local friction velocity, $u^*$, in the right column. The lines represent \protect\redline, case P2; \protect\blueReline, case P2I$_{Re}$; and \protect\darkgreenline, case P2O$_{Re}$. In the left column, solid lines represent the full velocity fluctuations and dashed lines represent the background-turbulence fluctuations. In the right column, only the background-turbulent fluctuations are portrayed. The solid black lines represent the smooth-wall simulations at $\Rey_\tau \approx 520$ and $1000$. The smooth-wall data at $Re_\tau \approx 1000$ is taken from \citet{Lee2015}.}}%
    	\label{fig:stats_resolved_gl_Re}%
\end{figure}

{The turbulent velocity fluctuations and Reynolds shear stresses for the simulations at $\Rey_\tau \approx 1000$, P2I$_{Re}$ and P2O$_{Re}$, are compared with those from case P2 in both global and local scaling in figure~\ref{fig:stats_resolved_gl_Re}.}

\begin{figure}
	\centering
  		\subfloat{%
%  		 \tikzsetnextfilename{urms_T_g}
%  		\input{images_Ts/u_T_g.tex}
  		\includegraphics[scale=1]{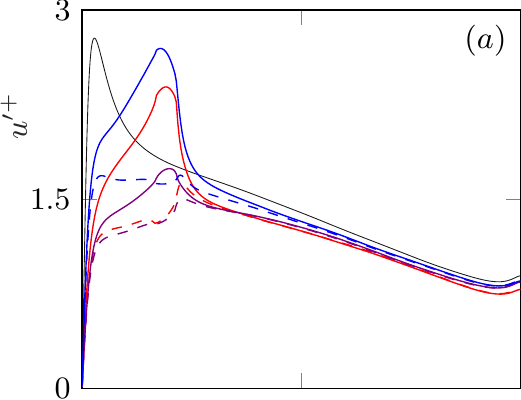}
  		}%
  		\hspace{3mm}\subfloat{%
%  		 \tikzsetnextfilename{urms_T_l}
%  		\input{images_PT_local/u_T_l.tex}
  		\includegraphics[scale=1]{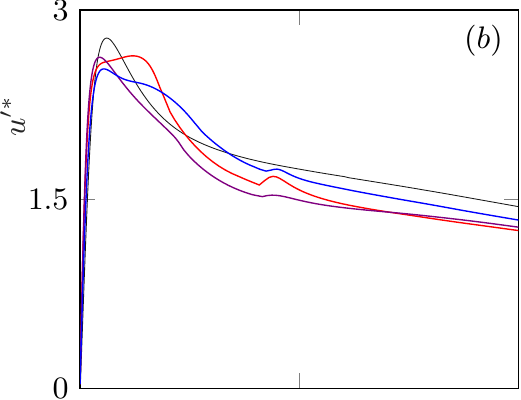}
  		}%
  		
  	    \vspace{0mm}\subfloat{%
%  		 \tikzsetnextfilename{vrms_T_g}
%  	       \input{images_Ts/v_T_g.tex}
  	       \includegraphics[scale=1]{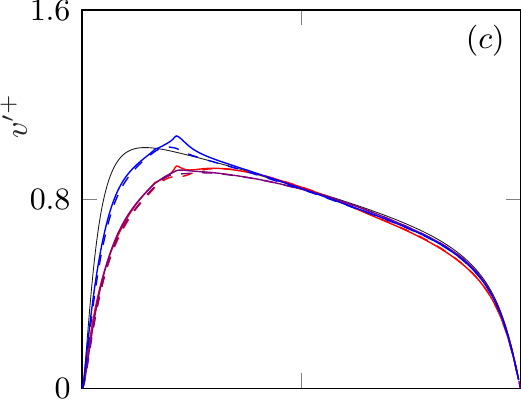}
  		}%
  		 \hspace{3mm}\subfloat{%
%  		 \tikzsetnextfilename{vrms_T_l}
%  		\input{images_PT_local/v_T_l.tex}
  		\includegraphics[scale=1]{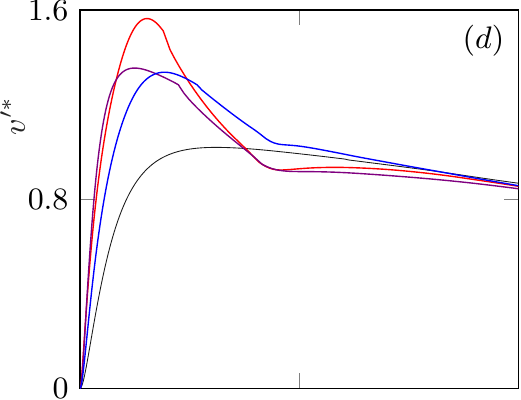}
  		}%
  		
  	  \vspace{0mm}\subfloat{%
% 		\tikzsetnextfilename{wrms_T_g}
% 			\input{images_Ts/w_T_g.tex}
 			\includegraphics[scale=1]{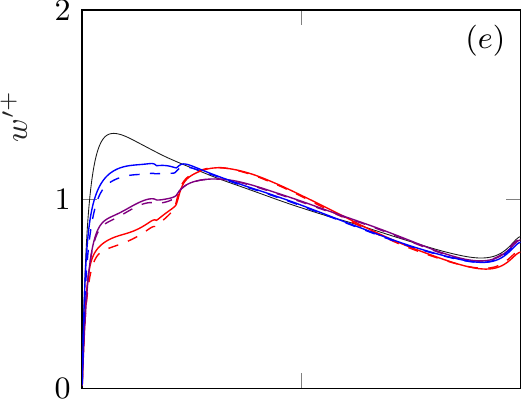}
 		}%
        \hspace{3mm}\subfloat{%
% 		\tikzsetnextfilename{wrms_T_l}
% 		\input{images_PT_local/w_T_l.tex}
 		\includegraphics[scale=1]{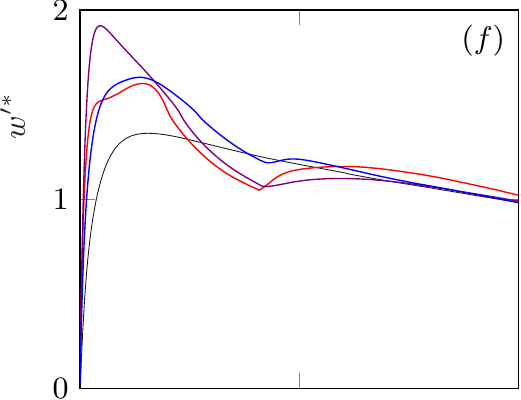}
 		}%		
 		
 		\vspace{3.5mm}\hspace{2.3mm}\subfloat{%
% 		\tikzsetnextfilename{uvrms_T_g}
% 			\input{images_Ts/Tau_T.tex}
 			\includegraphics[scale=1]{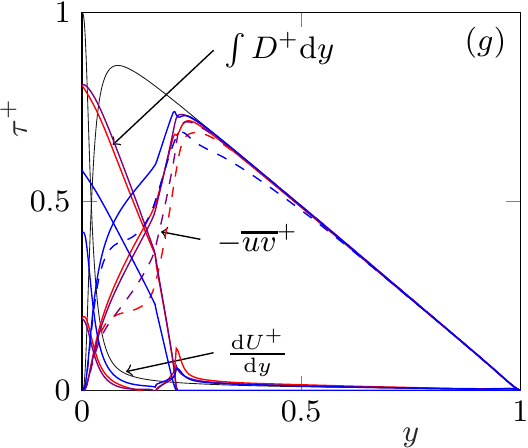}
 		}%
        \hspace{2.5mm}\subfloat{%
% 		\tikzsetnextfilename{uvrms_T_l}
% 		\input{images_PT_local/Re_stress_T_l.tex}
 		\includegraphics[scale=1]{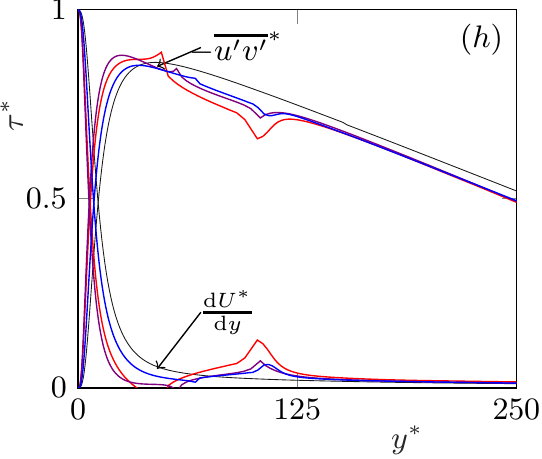}
 		}%
  	 \caption{Rms velocity fluctuations and shear stresses scaled with the global friction velocity, $u_\tau$,  in the left column and with the local friction velocity, $u^*$, in the right column. The lines represent \protect\redline, case T1; \protect\violetline, case TP1; and \protect\blueline, case T2. In the left column, solid lines represent the full velocity fluctuations and dashed lines represent the background-turbulence fluctuations. In the right column, only the background-turbulent fluctuations are portrayed. The black lines represent the smooth-wall case, S, for reference.}%
    	\label{fig:stats_resolved_gl_Ts}%
\end{figure}

{The turbulence statistics, in both global and local scaling, for the T-shaped canopies are portrayed in figure~\ref{fig:stats_resolved_gl_Ts}}. 

\begin{figure}
	\centering
  		 \subfloat{%
%  		 \tikzsetnextfilename{urms_PH_l}
%  		\hspace{-3mm}\input{images_posts/u_PH_l.tex}
  		\includegraphics[scale=1]{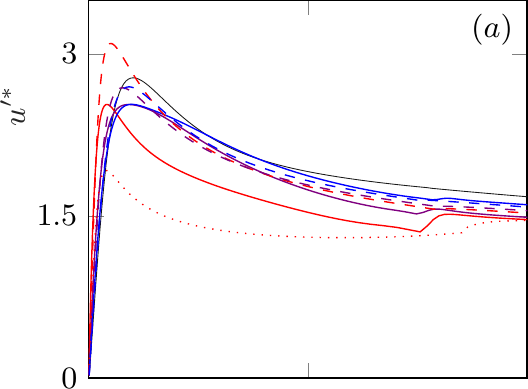}
  		}%
  		\hspace{5mm}\subfloat{%
%  		 \tikzsetnextfilename{urms_TH_l}
%  		\hspace{-7.5mm}\input{images_Ts/u_TH_l.tex}
  		\includegraphics[scale=1]{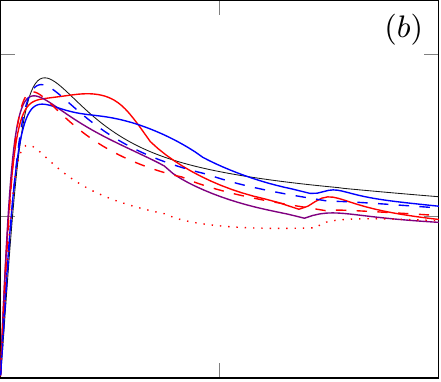}
  		}%
  		
   	\vspace{0mm}	\subfloat{%
%  		 \tikzsetnextfilename{vrms_PH_l}
%  		\hspace{-3mm}\input{images_posts/v_PH_l.tex}
  		\includegraphics[scale=1]{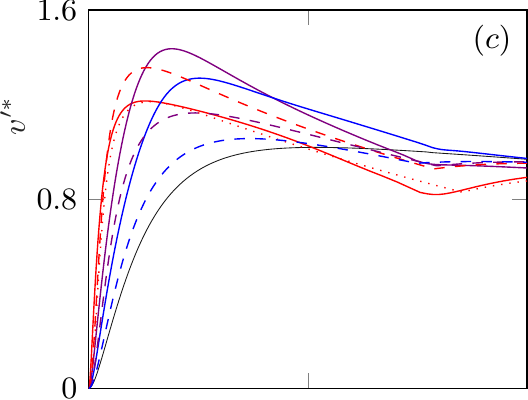}
  		}%
  		\hspace{5mm}\subfloat{%
%  		 \tikzsetnextfilename{vrms_TH_l}
%  		\hspace{-7.5mm}\input{images_Ts/v_TH_l.tex}
  		\includegraphics[scale=1]{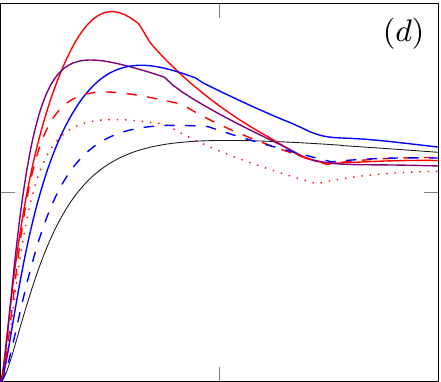}
  		}%
  	
        \vspace{0mm}\subfloat{%
% 		\tikzsetnextfilename{wrms_PH_l}
% 	    \hspace{-3mm}\input{images_posts/w_PH_l.tex}
 	    \includegraphics[scale=1]{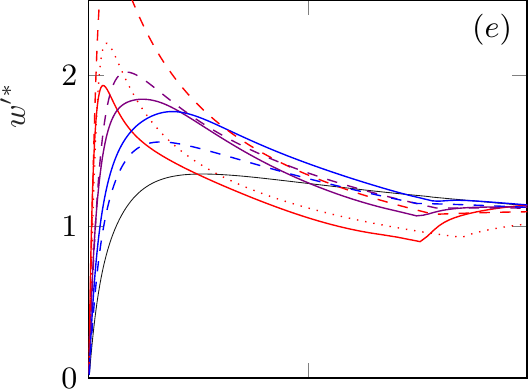}
 		}%	
 		\hspace{5mm}\subfloat{%
% 		\tikzsetnextfilename{wrms_TH_l}
% 		\hspace{-6mm}	\input{images_Ts/w_TH_l.tex}
 		\includegraphics[scale=1]{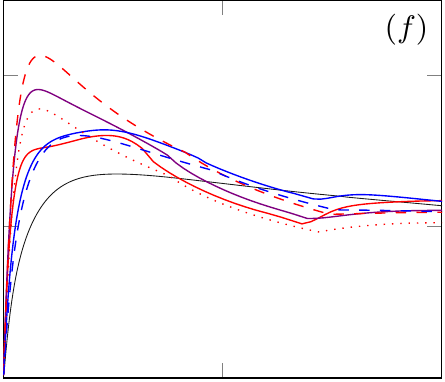}
 		}%		
 		 		 		 		
 		\vspace{3mm}\hspace{5.7mm}\subfloat{%
% 		\tikzsetnextfilename{uvrms_PH_l}
% 		\hspace{-0mm}	\input{images_posts/Re_stress_PH_l.tex}
 		\includegraphics[scale=1]{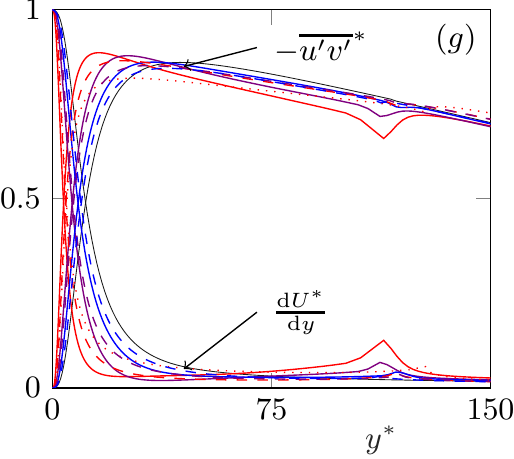}
 		}%
        \hspace{2.3mm}\subfloat{%
% 		\tikzsetnextfilename{uvrms_TH_l}
% 		\hspace{-8.5mm}	\input{images_Ts/Re_stress_TH_l.tex}
 		\includegraphics[scale=1,trim={0mm 0mm 0mm -0.7mm},clip]{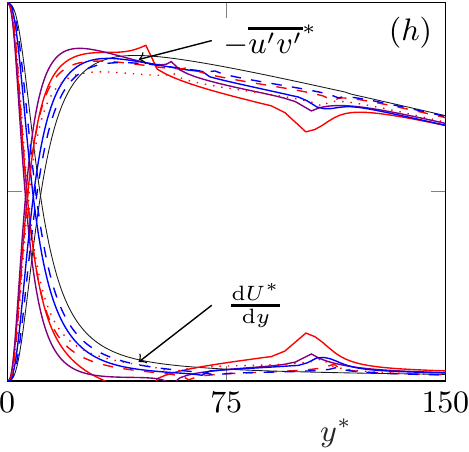}
 		}%
  	 \caption{Background-turbulence rms velocity fluctuations and shear stresses, scaled with the local friction velocity, $u^*$, of the resolved canopy and mean-only/homogeneous drag simulations. In the first column lines represent \protect\redline, case P0; \protect\reddashed, case P0-H0; \protect\reddotted, case P0-H; \protect\violetline, case P1; \protect\violetdashed, case P1-H0; \protect\blueline, case P2; and  \protect\bluedashed, case P2-H0. In the second column lines represent \protect\redline, case T1; \protect\reddashed, case T1-H0; \protect\reddotted, case T1-H; \protect\violetline, case TP1; \protect\blueline, case T2; and \protect\bluedashed, case T2-H0. The black lines represent case S.}%
	    	\label{fig:stats_models_l}	
\end{figure}

Figure~\ref{fig:stats_models_l} compares the background turbulence velocity fluctuations from the canopy resolving simulations with their corresponding mean-only or homogeneous drag simulations, in the proposed local scaling.

\bibliographystyle{jfm}
\bibliography{jfm_ref}
\end{document}